\documentclass[apj,numberedappendix]{emulateapj}
\usepackage{apjfonts}
\newcommand{\HI}{H{\sc~i}{ }}

\journalinfo{The Astrophysical Journal, in press}
\submitted{Received 2004 December 1; accepted 2005 February 4}
\shorttitle{Modeling Star Formation in Dwarf Spheroidals}
\shortauthors{Mashchenko, Couchman, and Sills}

\begin{document}

\title{Modeling star formation in dwarf spheroidal galaxies: a case for extended dark matter halos}

\author{Sergey Mashchenko, H. M. P. Couchman, and Alison Sills}

\affil{Department of Physics and Astronomy, McMaster University,
Hamilton, ON, L8S 4M1, Canada; syam,couchman,asills@physics.mcmaster.ca}

\begin{abstract}
We propose a simple model for the formation of dwarf spheroidal galaxies, in
which stars are assumed to have formed from isothermal gas in hydrostatic
equilibrium inside extended dark matter halos.  After expelling the leftover
gas, the stellar system undergoes a dynamical relaxation inside the dark matter
halo.  These models can adequately describe the observed properties of three
(Draco, Sculptor, and Carina) out of four Galactic dwarf spheroidal satellites
studied in this paper. We suggest that the fourth galaxy (Fornax), which cannot
be fitted well with our model, is observed all the way to its tidal radius.  Our
best fitting models have virial masses of $\sim 10^9$~M$_\odot$, halo
formation redshifts consistent with the age of oldest stars in these dwarfs, and
shallow inner dark matter density profiles (with slope $\gamma\sim -0.5\dots
0$).  The inferred temperature of gas is $\sim 10^4$~K.  In our model, the
``extratidal'' stars observed in the vicinity of some dwarf spheroidal galaxies
are gravitationally bound to the galaxies and are a part of the extended stellar
halos.  The inferred virial masses make Galactic dwarf spheroidals massive
enough to alleviate the ``missing satellites'' problem of $\Lambda$CDM
cosmologies.

\end{abstract}


\section{INTRODUCTION}

The popular $\Lambda$ cold dark matter ($\Lambda$CDM) cosmological concordance model has been very successful
in describing the properties of the universe on large and intermediate scales,
but has encountered numerous difficulties on smaller, galactic and sub-galactic scales
\citep[e.g.][]{tas03}. Among the most obvious discrepancies is the prediction
that there should be a factor of 50 more satellites orbiting in the
Milky Way halo, with the masses comparable and larger than the conventional
estimates for Galactic dwarf spheroidals (dSphs), than is actually observed ---
the so-called ``missing satellites'' problem \citep{moo99}. \citet{sto02} and
\citet{hay03} gave a possible solution to this problem. They argued that the
properties of Galactic dSphs are consistent with them being embedded in very
extended dark matter (DM) halos, with total masses comparable to the masses of the largest
substructure predicted to populate the Milky Way halo by $\Lambda$CDM
cosmologies.

There have also been cosmology-independent arguments in favor of much more
massive dSphs than the conventional $\sim 10^7$~M$_\odot$ estimates
 based on the ``mass follows light'' assumption \citep{mat98}. 
\citet{lok02} modeled the line-of-sight velocity dispersion profiles for Fornax and Draco
under the assumption that the anisotropy parameter $\beta\equiv
1-(\sigma_t/\sigma_r)^2$ is a constant, and derived the total masses for these
galaxies of $\sim (1-4)\times 10^9$~M$_\odot$. (Here $\sigma_t$ and $\sigma_r$
are the tangential and radial velocity dispersion, respectively.)  From fitting
a two parameter family of spherical models of \citet{wil02} to the observed
line-of-sight velocity dispersion profile in Draco, \citet{kle02} concluded that
the mass-to-light ratio in this dwarf increases outwards. \citet{ode01} used
large-field multicolor photometry of an area of 27 square degrees in the
vicinity of Draco to show that this galaxy has very regular stellar isodensity
contours down to their sensitivity limit (0.003 of the central surface
brightness), with no signs of tidal interaction with the Milky Way.  This
implies that Draco has more DM in its outskirts than in its center and a
practically unconstrained upper limit on total mass. \citet{bur97} and
\citet*{MCB04} proposed two different models (supernovae Ia heating of the ISM and 
radiation harassment by the ionizing photons from the host galaxy,
respectively) to explain complex star formation history of some dSphs.  The
important ingredient of both scenarios is the requirement for dSphs to be
very massive ($\gtrsim 10^9$~M$\odot$) so that the fully ionized ISM would
not escape to intergalactic space.

If it is indeed the case that many Galactic dSphs have massive,
$10^9-10^{10}$~M$_\odot$ DM halos (which would translate into their apparent
tidal radii being $\gtrsim 5$ angular degrees in the sky), then the observed
distribution of stars in these galaxies should have been entirely shaped by
internal processes, and not by the tidal field of the Milky Way. The current
stellar density in Galactic dSphs is extremely low (they are DM dominated even
at the center), with the estimated dynamical relaxation time for stars being
much larger than the Hubble time (assuming that the DM halo is sufficiently
smooth at the center). It is reasonable to assume then that the distribution of
stars in dSphs has stayed unchanged since the initial star formation epoch.

\citet{lak90} noted that the core line-of-sight velocity dispersion in Draco
and Ursa Minor ($\sim 10$~km~s$^{-1}$) is of the same order as the sound speed in
a diffuse cooling cloud of primordial composition. He argued that a simple model
of stars forming from isothermal gas in hydrostatic equilibrium in a DM halo
with the Plummer density profile provides a reasonably good description of the
observed properties of these two dSphs. His conclusion was that the apparent
cutoff radius in the distribution of stars in Draco and Ursa Minor has nothing
to do with galactic tides.

Here we propose a simple star formation model for dSphs. We assume that stars
formed from isothermal self-gravitating gas in hydrostatic equilibrium
inside an extended DM halo. We simulate the relaxation of the stellar cluster
after the leftover gas is expelled using an $N$-body code. We significantly
improve upon the model of
\citet{lak90} by (1) considering self-gravity of gas and stars, (2) allowing for
dynamical relaxation of stars after expelling the leftover gas, (3) introducing
a lower density cutoff for star formation, and (4) using properly 
normalized $\Lambda$CDM DM halos.  We show that our models are consistent with the existing
observations on three out of four dSphs studied in this paper, if one assumes
that these dwarfs have very massive ($\sim 10^9$~M$_\odot$) DM halos.

The paper is organized as follows. In Section~\ref{model} we describe our model.
In Section~\ref{comparison} we present our algorithm for finding best fitting
models. We present our results for four dSphs in Section~\ref{results}.  We
discuss the results in Section~\ref{discussion}, and give our conclusions in
Section~\ref{conclusions}.

\section{MODEL}
\label{model}

\subsection{Overview and Assumptions}

We assume that initially gas was isothermal and in hydrostatic equilibrium in
the static potential of the DM halo. Stars are assumed to have formed from gas
instantaneously with a certain efficiency $\xi$ out to a radius where the gas
density drops below some critical value. The velocity dispersion of newly born
stars is assumed to be equal to the sound speed in the gas.  After the leftover gas
is expelled instantaneously by the combined action of ionizing radiation,
stellar winds, and supernova explosions, the stellar cluster is relaxed over a
few crossing times through violent relaxation and/or phase mixing. We simulate
this process numerically using an $N$-body code.

As we will show in this paper, in our model the stellar cluster does not expand
much after the star burst, resulting in the final central velocity dispersion
being only $\sim 20$\% smaller than the initial value. The central line-of-sight
velocity dispersion for dSphs is in the narrow interval $8.5\pm 2$~km~s$^{-1}$
\citep{mat98}. It is remarkable that this parameter is almost constant for all
Galactic dSphs. It is also comparable to the sound speed in the warm neutral
medium, $6-7.5$~km~s$^{-1}$, in the ISM of the Milky Way (\citealt{wol95}; we used their
``standard'' model and assumed that the mean molecular weight is $14/11 m_p$,
where $m_p$ is the mass of the proton), and ISM velocity dispersion of $\sim
9\pm 2$~km~s$^{-1}$ in dwarf irregular galaxies
\citep{mat98}. The vertical velocity dispersion of young stars in the Galactic
disk, $6\pm 3$~km~s$^{-1}$ \citep{hay97}, is also in the same range, giving
support to our model assumption that the velocity dispersion of newly born stars
is comparable to the sound speed in the star-forming gas.

To justify our assumption of the gas being isothermal, we should mention that in
all of our models the relaxed stellar clusters are almost isothermal (in terms
of total stellar velocity dispersion). Galactic dSphs are known to have roughly
isothermal line-of-sight velocity dispersion profiles in their central regions
\citep{mat97,kle02,wil04}. We can also reproduce the observed decline in
the line-of-sight velocity dispersion in the outskirts of some dSphs
\citep{kle04}, which in our model is caused by a fraction of stars which move
beyond the initial stellar cluster radius during the dynamical relaxation
phase, and as a result have strong radial anisotropy.

Because the DM halo potential in our model is static, neither adiabatic
contraction of the central part of DM halo due to presence of baryons nor
adiabatic expansion after the removal of the remaining gas in a post-starburst
system are modeled. Static potential models presented in this paper took around
500 CPU-days to compute; live DM halo simulations would take orders of magnitude
longer to run, which is obviously not feasible. From both observational and
theoretical points of view, the situation with inner density profiles of DM
halos is less than clear. Observations produce less steep inner density slopes
for galactic DM halos ($\gamma=-1\dots 0$) than the predictions of cosmological
CDM models ($\gamma=-1.5\dots -1$). Though a few mechanisms [such as warm DM
\citep{bod01}, recoiling black holes \citep{boy04}, and dynamic heating due to
presence of dense gas clumps \citep{elz01}] were proposed to bring theory in
agreement with observations, none of these explanations has received yet wide
acceptance. Given the uncertainty in the inner density slope for galactic DM
halos, we believe that by considering two different types of static DM halos,
with flat core ($\gamma=0$) and with $\gamma=-1$ cusp, we alleviated some
inadequacies resulting from the use of static DM potential.

\subsection{Initial Gas Distribution}

Hydrostatic equilibrium of self-gravitating gas inside the static potential of a
spherical DM halo can be described using the following expression:

\begin{equation}
\label{hydstat}
\frac1{\rho_g}\frac{dP_g}{dr} = -\frac{d\Phi_{\rm tot}}{dr}.
\end{equation}

\noindent Here $P_g$ and $\rho_g$ are the pressure and the density of gas, and
$\Phi_{\rm tot}$ is the total gravitational potential with contributions from
both gas and DM. The radial gradient of gravitational potential in
equation~(\ref{hydstat}) can be calculated as $d\Phi_{\rm
tot}/dr=G[M(r)+m_g(r)]/r^2$, where $M(r)$ and $m_g(r)$ are the enclosed masses
of DM and gas, respectively, and $G$ is the gravitational constant. To close
equation~(\ref{hydstat}), one has to write Poisson equation for the gas
component:

\begin{equation}
\label{poi}
\frac1{r^2}\frac{d}{dr}\left(r^2\frac{d\varphi_g}{dr}\right)=4\pi G \rho_g.
\end{equation}

\noindent Here $\varphi_g(r)$ is the gas contribution to the total potential.

We consider two types of DM halo density profiles: flat-core \citet{bur95} halos and
Navarro-Frenk-White (NFW) halos which have a central density cusp with $\gamma=-1$
\citep*{NFW97}. The density profiles for these halos are as follows:

\begin{eqnarray}
\rho(r)& = &\frac{\rho_{0}}{(1+r/r_s)\,[1+(r/r_s)^2]}\quad\mbox{(Burkert),}\\
\rho(r)& =& \frac{\rho_{0}}{r/r_s\, (1+r/r_s)^2}\quad\mbox{(NFW).}
\end{eqnarray}

\noindent Here $r_s$ and $\rho_0$ are the scaling radius and density for DM halos.
(In Burkert halos, $\rho_0$ corresponds to the central DM density; in NFW halos,
which have divergent central density, $\rho=\rho_0$ at $\sim 0.5r_s$.)

For the following analysis, it is convenient to switch to new dimensionless
variables: $x\equiv r/r_s$ for radial distance, $\rho'\equiv \rho/\rho_0$ for
density, $M'\equiv M/(4\pi r_s^3
\rho_0)$ for mass,  $v'\equiv v/(4\pi G r_s^2 \rho_0)^{1/2}$ for velocity,
and $t'=t (4\pi G \rho_0)^{1/2}$ for time. In these new variables, the enclosed
DM mass for the two types of halos can be written as follows:

\begin{equation}
\label{Ms}
M'(x)=\cases{
(1/2)[\ln(1+x)+(1/2)\ln(1+x^2)-\cr
\quad -\arctan x] & \mbox{(Burkert),}\cr
\ln(1+x)-x/(1+x) & \mbox{(NFW)}.\cr
}
\end{equation}

We assume that gas is isothermal: $P_g=\rho_g c_g^2$ (here $c_g$ is the sound
speed in gas, which is assumed to be constant). Equations~(\ref{hydstat}) and
(\ref{poi}) then can be rewritten as

\begin{equation}
\label{ode1}
\frac{d\rho'_g}{dx}=-\frac1{c_g'^2}\frac{\rho'_g}{x^2}[M'(x)+m'_g],
\end{equation}

\noindent and

\begin{equation}
\label{ode2}
\frac{dm'_g}{dx}=x^2\rho'_g.
\end{equation}

Equations~(\ref{Ms}--\ref{ode2}) form a system of two ordinary differential
equations for two variables, $\rho'_g(x)$ and $m'_g(x)$, which has to be solved
numerically. The boundary conditions are $\rho'_g(0)=\rho'_{g,0}$ (a free
parameter) and $m'_g(0)=0$.

\subsection{Star Formation and Dynamical Relaxation of Stars}
\label{SFDR}

We adopt a simple star formation model.  We assume that a certain mass fraction
$\xi$ of gas is turned instantaneously into stars everywhere inside the halo
where the dimensionless gas density is above a certain threshold value $\lambda$
(in units of the central gas density $\rho'_{g,0}$). In other words, star
formation efficiency is assumed to be $\xi$ (a constant) for $\rho_g'>\lambda
\rho'_{g,0}$, and zero otherwise. The parameter $\lambda$ was introduced to
avoid the star formation occuring at unphysically low densities.  The newly born
stars are assumed to have an isotropic distribution of velocity vectors, with
the one-dimensional velocity dispersion equal to the sound speed in gas $c_g'$.

After the starburst, all the remaining gas is assumed to be removed from the
system instantaneously by the combined action of ionizing radiation, stellar
winds, and supernovae explosions. In reality, a fraction of this gas can be
later reaccreted by the halo leading to subsequent episodes of star
formation. For simplicity, here we consider all stars in a dwarf spheroidal
galaxy to be formed in a single starburst.

It is easy to see that models with $\xi=1$ and $\lambda=0$ will remain in
equilibrium after all gas is turned into stars. Indeed, our hydrostatic gas
equilibrium equation~(\ref{hydstat}) is identical to the isotropic stellar Jeans
equation if $\rho_*=\rho_g$ and $\sigma_*=c_g$. (Here $\rho_*$ and $\sigma_*$
are density and one-dimensional velocity dispersion of stars.)

For the cases with $\xi<1$ and/or $\lambda>0$, the initial configuration of
stars after the gas removal is not an equilibrium one: stars are dynamically
hot, and will expand before reaching an equilibrium state. This highly
non-linear process involves phase mixing and/or violent relaxation. Consequently,
we had to resort to numerical $N$-body simulations to derive the final equilibrium
configuration of our stellar clusters. To run the simulations, we used the
parallel version of the popular multistepping tree code Gadget \citep{spr01}.

\subsection{Numerical Scheme}
\label{numer}

Overall, there are four free parameters in our model which can influence
properties of relaxed stellar clusters: the sound speed in gas in units
of the halo scaling velocity $c_g'$, the central density of gas in units of
the halo scaling density $\rho_{g,0}'$, the star formation efficiency $\xi$, and
the minimum gas density $\lambda$ for star formation to take place in units
of the central gas density.

The parameter $c_g'$ can be considered as a measure of how hot gas is in
relation to the virial temperature of the halo. The nuance is that in our model
the gas temperature is measured in terms of ``scaling temperature'' which can be
defined as $T_s=\mu V_s^2/(2k)$, whereas the usual definition for the virial
temperature of a halo is $T_{\rm vir}=\mu V_{\rm vir}^2/(2k)$. (Here $\mu$ is
the mean molecular weight of the gas, $k$ is the Boltzmann constant, $V_s$ and
$V_{\rm vir}$ is the halo circular speed at the scaling radius $r_s$ and the
virial radius $r_{\rm vir}$, respectively.) The connection between $T_s$ and
$T_{\rm vir}$ depends on halo density profile (Burkert or NFW) and concentration
$c= r_{\rm vir}/r_s$.  For example, for both Burkert and NFW halos with
$c=3.5\dots 10$, gas in our models is at the virial temperature when
$c_g'=0.46\dots 0.39$.

We consider three values of the parameter $\lambda$: 0 (no star formation
cutoff), 0.1, and 0.3. We cover a wide range of values for the remaining three
parameters: $c_g'=0.01\dots 0.48$, $\rho_{g,0}'=0.1\dots 10^5$, and
$\xi=10^{-3}\dots 1$.

To generate an initial particle distribution for the $N$-body simulations, we use the
enclosed gas mass function $m'_g(x)$ which is derived numerically by solving
equations~(\ref{Ms}--\ref{ode2}). First, we find the star formation cutoff
radius $x_\lambda$ such as $\rho_g'(x_\lambda)=\lambda\rho'_{g,0}$. (For the case of
$\lambda=0$, $x_\lambda$ is set to be equal to the largest considered radius
$x_{\rm max}=10^3$). Next, the integral probability function $P(x)$ is
constructed as $P(x)=m'_g(x)/m'_g(x_\lambda)$. For each $N$-body particle, the
radial distance $x$ is obtained by generating a uniform random number $y=[0\dots
1]$ and then solving numerically the equation $P(x)=y$. Finally, to obtain three
components of the radius and velocity vectors we assume that they are randomly
oriented in space, which is true for a spherically symmetric system with
isotropic distribution of velocity vectors. We assumed Maxwellian distribution
of velocities, with the one-dimensional velocity dispersion equal to the sound
speed in the gas $c_g'$.

All our models were simulated with $N=10^4$ equal mass stellar particles. This
number is large enough to allow for accurate surface brightness profile
calculations over a large range of radial distances and to avoid significant
spurious dynamical evolution caused by close encounters between particles.  At
the same time, $N$ is small enough to make it feasible to simulate hundreds of
models. The masses of individual particles are equal to $m_*/N$, where $m_*=\xi
m'_g(x_\lambda)$ is the total stellar mass.

The value of the gravitational softening length $\varepsilon$ was calculated as $0.77 x_h
N^{-1/3}$ \citep{hay03}, where the initial stellar half-mass radius $x_h$ was
derived by solving the equation $P(x_h)=1/2$. The values of the Gadget
parameters which control accuracy of integration were the same as in
\citet{mas05a,mas05b}, where a total energy conservation of better than 2\% was
achieved when simulating a globular cluster for 6000 crossing times at the half-mass
radius. In particular, the individual time steps in the simulations were equal
to $(2\eta \varepsilon/a)^{1/2}$, with the accuracy parameter
$\eta=0.0025$. (Here $a$ is the acceleration of a particle.) The parameter
controlling the force calculation accuracy (ErrTolForceAcc in Gadget notation)
was set to 0.01.

In the $N$-body simulations, the DM halo is represented by a static gravitational
potential, which is identical to the one used for solving
equations~(\ref{Ms}--\ref{ode2}).  The important point to make is that our DM
halos are not truncated at the virial radius --- neither in solving
equations~(\ref{Ms}--\ref{ode2}) nor in $N$-body simulations. We did the
consistency check (that the observed size of the galaxy is smaller than the
inferred virial and tidal radii) after fitting the models to observations. 

We ran our $N$-body models for the interval of time $t'=370$ (in dimensionless
units), with 200 intermediate snapshots per model. This time interval is much
larger than the crossing time $t'_{\rm cross}=1/c_g'$ at the scaling radius $r_s$,
which is $\leqslant 100$ for all our models. 

To cover well the 4-dimensional free parameter space, we had to run a
significant number of models (around 700). To facilitate the process, we
designed a pipeline for automatic generation of observable properties of the
models for given DM halo profile (Burkert or NFW) and given values of the four
initial free parameters $c_g'$, $\rho_{g,0}'$, $\xi$, and $\lambda$. The
pipeline consists of the following steps. 

\begin{enumerate}
\item System of differential
equations~(\ref{Ms}--\ref{ode2}) is solved numerically, producing isothermal
hydrostatic gas distribution in the potential of the DM halo. 

\item Using the
procedure described above in this section, the initial distribution of stellar
particles is generated. 

\item An $N$-body code is run
to simulate the stellar cluster relaxation in the static DM halo potential.

\item For last $N_S=100$ snapshots
(corresponding to $t'=185\dots 370$), both time-averaged values and standard
deviations from mean are calculated for surface brightness $\Sigma$ in different
radial bins. To calculate $\Sigma$, we use the projection method described in
Appendix~B of \citet{mas05a}.  The noisier innermost and outermost parts of the
surface brightness profile are cut off at the radii where the standard deviation
from mean becomes larger than $\Delta=0.2$~dex. In this fashion, we ensure that
the inner part of the $\Sigma$ profile has converged to better than
$\Delta/N_S^{1/2}=0.02$~dex (because of very short crossing time near the halo
center resulting in virtually uncorrelated surface brightness values in
different snapshots), and that the outer part of the profile (where a strong
correlation is expected) has converged to $0.02\dots 0.2$~dex. We repeat this
procedure for a smaller number of snapshots $N_S=50$ (corresponding to
$t'\sim 280\dots 370$), and choose the profile which has converged over a larger
range of radial distances.

\item We calculate a line-of-sight velocity
dispersion $\sigma_0$ [using a procedure delineated in Appendix~B of
\citet{mas05a}] averaged over the core region of the galaxy (within the
half-brightness radius) and time-averaged over the last $N_S$ snapshots.
\end{enumerate}

As a test, we compared the radial stellar density profiles in our evolved models
with $\xi=1$ and $\lambda=0$ with the corresponding initial gas density profiles
(they should be identical --- see \S~\ref{SFDR}). In most of the models, the
profiles are found to be identical within the measurements errors. In a few high
density models, with $\rho'_{g,0}=10^3$ and $10^5$, we see spurious flattening
of the inner stellar density profile due to interaction between stellar
particles. This effect is found to be negligible at the radii larger than the
softening length $\varepsilon$.

\subsection{General Properties of the Models}
\label{props}

Surface brightness profiles in our relaxed models exhibit a variety of shapes
--- from simple cases of monotonously increasing slope, to profiles with one or
more inflection points.  The important fact is that in most cases we observe a
flat core plus extended halo (with different slopes) in the surface brightness
profiles for both Burkert and NFW static halos --- somewhat like profiles of
dSphs. An obvious central stellar cusp is seen only in the models with strongly
self-gravitating gas: $\rho_{g,0}'\gtrsim 10^3$ (for both Burkert and NFW halos).

Our relaxed stellar models are approximately isothermal (in terms of total
three-dimensional velocity dispersion).  For models with non-zero star formation
lower density cutoff, $\lambda$, the line-of-sight velocity dispersion profiles
are often sharply declining at large radii.  For such models, the outer stellar
halo is strongly radially anisotropic.

\begin{figure}
\plotone{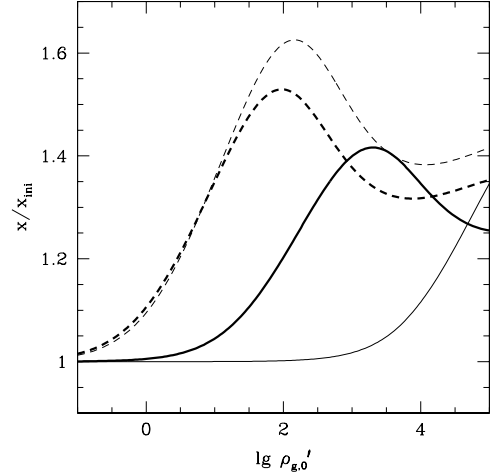}
\caption {Dependence of the estimated radial expansion factor $x/x_{\rm ini}$ on
initial central gas density $\rho_{g,0}'$ (in the limit $\xi\rightarrow
0$). Thick and thin lines correspond to models with $c_g'=0.16$ and 0.01,
respectively. Solid (dashed) lines correspond to NFW (Burkert) halos. In all the
models $\lambda=0$.
\label{fig1} }
\end{figure}

\begin{figure*}
\plottwo{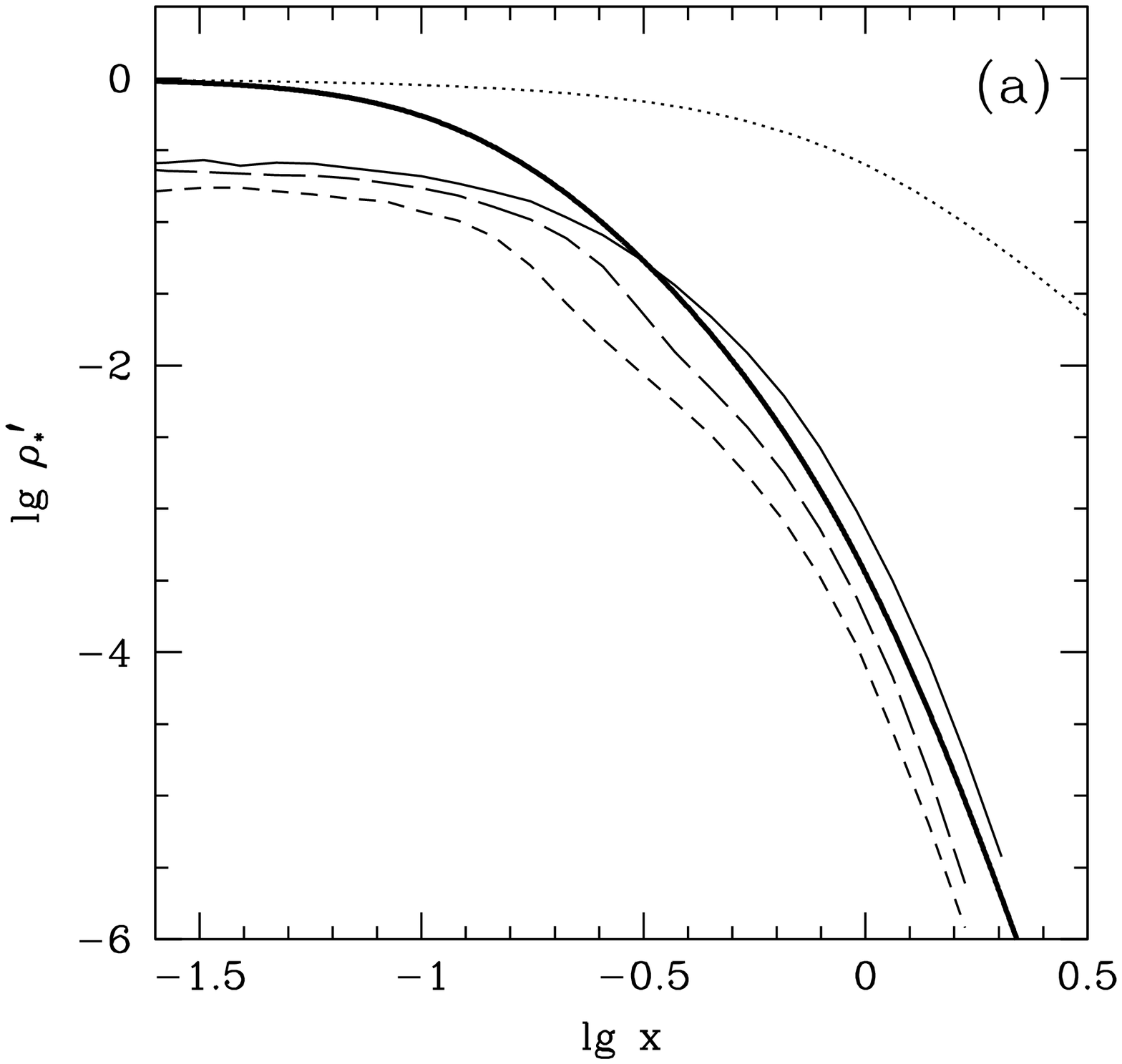}{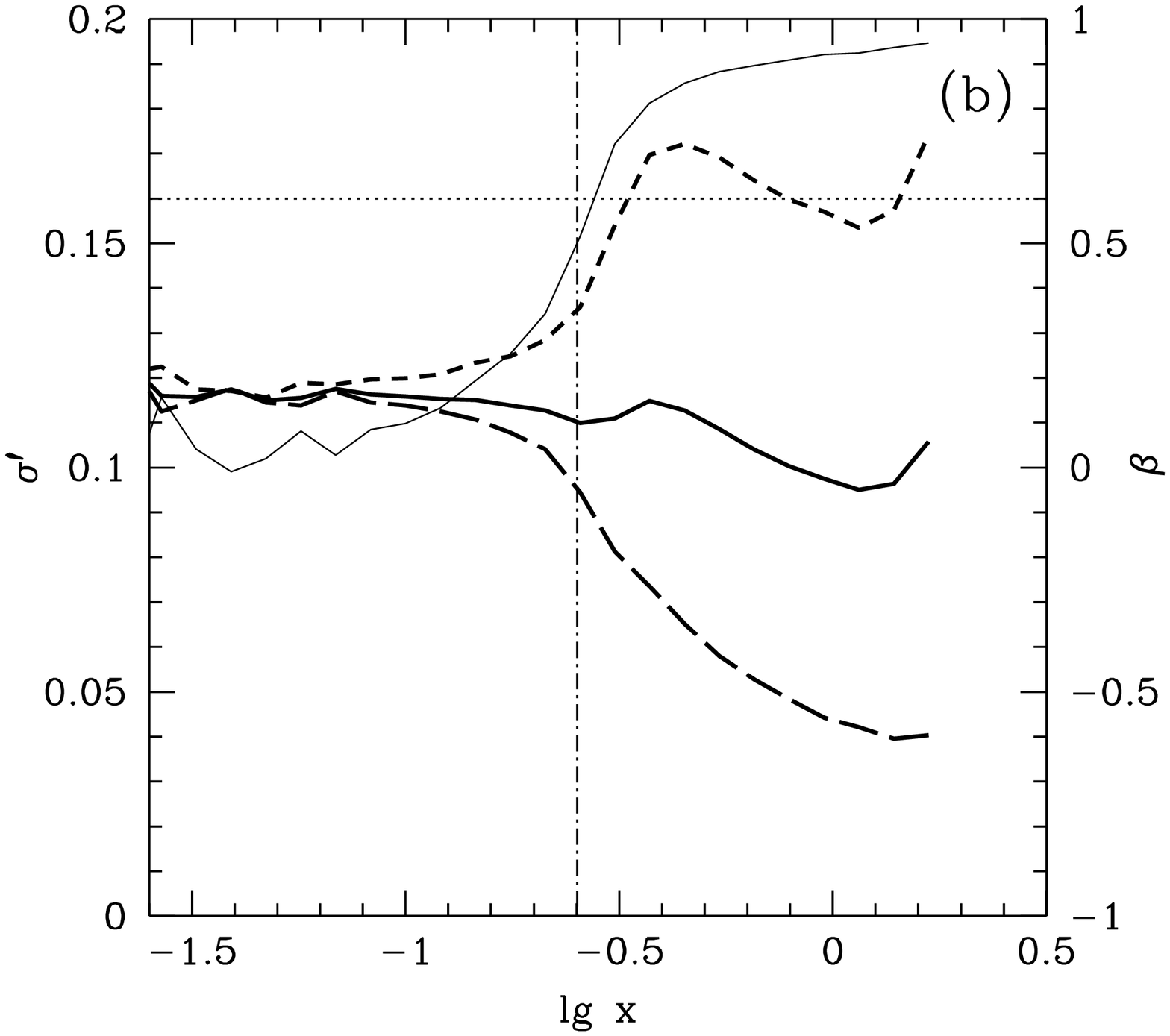}
\caption {Radial profiles for Burkert models with $c_g'=0.16$, $\rho_{g,0}'=10$, and
$\xi=0.1$. (a) Stellar density profiles. Thick solid line shows the initial
non-equilibrium profile. Profiles for relaxed models with $\lambda=0$, 0.1, and
0.3 are shown as thin lines (solid, long-dashed, and short-dashed,
respectively).  The dotted line shows the DM density profile. (b) Stellar
one-dimensional velocity dispersion $\sigma'$ (left ordinate) and anisotropy
$\beta=1-(\sigma'_t/\sigma'_r)^2$ (right ordinate) as a function of radial
distance $x$ for model with $\lambda=0.1$. Thick solid, short-dashed, and
long-dashed lines correspond to total dispersion $\sigma'$, radial dispersion
$\sigma'_r$, and tangential dispersion $\sigma'_t$ for the relaxed
model. Horizontal dotted line shows the initial non-equilibrium value of velocity
dispersion. Vertical dash-dotted line marks the cutoff radius $x_\lambda$.
Thin solid line shows radial $\beta$ profile for the relaxed model.
\label{fig2} }
\end{figure*}

On average, our relaxing stellar clusters do not expand much, and as a
consequence do not cool down much adiabatically. For all our models, the ratio
of initial (non-equilibrium) core line-of-sight velocity dispersion $\sigma_{\rm
ini,0}$ to its final, relaxed value $\sigma_0$ varies between $-0.01$ and
0.43~dex, with a small average value of $0.10\pm 0.07$~dex.  To explain this
model property, let us consider the extreme case of $\xi\rightarrow 0$ resulting
in the largest possible expansion factor. Initially, stars are born on
equilibrium orbits (due to the fact that our hydrostatic equilibrium equation is
identical to the isotropic stellar Jeans equation -- see Section~\ref{SFDR}).
An instantaneous removal of virtually all baryons after the starburst does not
change individual orbital angular momenta of the stars, resulting in the
following adiabatic law:

\begin{equation}
\label{adiabata}
\frac{\sigma'}{\sigma_{\rm ini}'}=\frac{x_{\rm ini}}{x}.
\end{equation}

\noindent Here $x$ is the baryonic half-mass radius and $\sigma'$ is stellar 
velocity dispersion. Suffix ``ini'' stands for the initial, non-equilibrium
values. The virial equation for the relaxed cluster in case of $\xi\rightarrow
0$ is

\begin{equation}
\label{virial}
\left(\frac{\sigma'}{\sigma_{\rm ini}'}\right)^2\simeq \frac{x_{\rm ini}}{x}\frac{M'(x)}{M'(x_{\rm ini})+m'_g/2},
\end{equation}

\noindent where the gravitational radius is assumed to be equal to
the half-mass radius. We estimate the largest expansion factor $x/x_{\rm ini}$ by
combining equations~(\ref{adiabata}) and (\ref{virial}):

\begin{equation}
\frac{x}{x_{\rm ini}}\simeq \frac{M'(x_{\rm
ini})+m'_g/2}{M'(x)}.
\end{equation}

\noindent  Assuming that most of the gas is concentrated within the scaling
radius of the DM halo (which is the case for almost all of our models), $M'(x)$
can be approximated as $M'(x)\propto x^\alpha$, where $\alpha=2$ for NFW and 3
for Burkert halos. In this case, the expansion factor can be written as

\begin{equation}
\frac{x}{x_{\rm ini}}\simeq \left[1+\frac{m'_g}{2M'(x_{\rm ini})}\right]^{1/(\alpha+1)}.
\end{equation}

\noindent   It happens,
that for isothermal gas in hydrostatic equilibrium inside a static DM halo, the
estimated expansion factor $x/x_{\rm ini}$ is never larger than $\sim 1.7$ for
the whole range of considered $c_g'$ and $\rho_{g,0}'$, and for both NFW and
Burkert halo density profiles (see Figure~\ref{fig1}). For less extreme cases
(with larger $\xi$) we expect the expansion factor to be even closer to unity.

To illustrate the impact of a lower density cutoff for star formation $\lambda$
on our models, we show in Figure~\ref{fig2} a few radial profiles for Burkert
models with $c_g'=0.16$, $\rho_{g,0}'=10$, and $\xi=0.1$. From the panel (a) of
this figure one can see that non-zero values of $\lambda$ both change global
properties of the relaxed stellar clusters (such as central density and size)
and modify the shape of the density profile. In panel (b) of Figure~\ref{fig2}
the radial velocity dispersion profiles are shown for the model with
$\lambda=0.1$. (This model is close to some of the best fitting models for
Galactic dSphs --- see Section~\ref{results}.) As you can see, stars in the
relaxed cluster show strong radial anisotropy (with the anisotropy parameter
$\beta\rightarrow 1$) outside of the initial radius of the cluster $x_\lambda$ (the
vertical dash-dotted line). The total velocity dispersion (thick solid line)
does not change much with radius. The transition from an almost isotropic to
a strongly radially anisotropic distribution of velocities is very
sharp in this model. It is interesting to note that in the outskirts of the
cluster the radial velocity dispersion (short-dashed thick line) is comparable
to the initial velocity dispersion (horizontal dotted line).

\section{COMPARISON WITH OBSERVATIONS}
\label{comparison}

\subsection{General Remarks}
\label{par0}

\begin{table}
\caption{Observational data on dwarf spheroidals\label{tab1}} 
\begin{tabular}{lccccc}
\tableline
Name     & $D$ & $\Sigma_0$        & $\sigma_{\rm obs,0}$    & $\Upsilon$               & $\chi^2_{\rm King}$ \\
         & kpc & mag~arcsec$^{-2}$ & km~s$^{-1}$   & M$_\odot$~L$_\odot^{-1}$ & \\
\tableline
Draco    & 82  & 25.3              & $9.5\pm 1.6$  & 1.32                     & 13.2\tablenotemark{a}, 147.0\tablenotemark{b}\\
Sculptor & 79  & 23.7              & $6.6\pm 0.7$  & 1.23                     & 296.1\\
Carina   & 101 & 25.5              & $6.8\pm 1.6$  & 0.85                     & 28.7\\
Fornax   & 138 & 23.4              & $10.5\pm 1.5$ & 0.94                     & 304.9\\
\tableline
\end{tabular}
\tablecomments{Here $\chi^2_{\rm King}$ is the $\chi^2$ value for the best fitting theoretical King model.}
\tablenotetext{a}{For the dataset Draco1 [from \citet{ode01}].}
\tablenotetext{b}{For the dataset Draco2 [from \citet{wil04}].}
\end{table}

We use two types of observational data on Galactic dwarf spheroidal galaxies in
our analysis: star number density profiles (with corresponding one-sigma
uncertainties) and core line-of-sight velocity dispersion $\sigma_{\rm obs,0}$ [taken from
\citet{mat98}, and listed in our Table~\ref{tab1}]. The star number density
profiles for four dwarf spheroidals analyzed in this paper (Draco, Sculptor,
Carina, and Fornax) are taken from \citet{ode01}, \citet{wil04}, and
\citet{wal03}. For our dwarf galaxies, we adopt distance $D$ and central surface 
brightness $\Sigma_0$ values from \citet{mat98}, and baryon mass-to-light ratio
$\Upsilon$ from \citet[their Table~6; we averaged $\Upsilon$ values for Salpeter
and ``Composite'' initial mass functions]{mov98}. These three parameters ($D$,
$\Sigma_0$, and $\Upsilon$; listed in Table~\ref{tab1}) are used to convert the
observed star number density profiles into stellar surface density profiles
$\Sigma(r)$ (in M$_\odot$~pc$^{-2}$ units). To perform the conversion, we
estimated the central star number density in the observed profiles by averaging
values for a few central radial bins, well within the galactic core.

Due to the time-consuming nature of the simulations, high dimensionality of our free
parameter space, and some degree of noisiness in the results of our simulations,
we do not attempt to find a model which is the best match to
observations. Instead, for each galaxy we try to find at least one
(typically more) models which would simultaneously satisfy the following
conditions: (1) Surface brightness profile should be very similar to the
observed one (quantified in \S~\ref{par1}).  (2) Core (within the half-brightness
radius) line-of-sight velocity dispersion $\sigma_0$ should be equal to the
observed value $\sigma_{\rm obs,0}$ (Table~\ref{tab1}) within the measurement
errors.  (3) The values of the DM halo scaling mass $M_s=4\pi r_s^3\rho_0$ and
scaling radius $r_s$, which are inferred after fitting the model surface
brightness profile to the observed one, should correspond to realistic halos
predicted by cosmological $\Lambda$CDM simulations (quantified in
\S~\ref{par3}).

\subsection{$\chi^2$ Fitting of Surface Brightness Profiles}
\label{par1}

To quantify the first of the selection criteria in \S~\ref{par0}, we measure
weighted $\chi^2$ between the model and observed surface brightness profiles (in
linear space).  We minimize $\chi^2$ numerically as a function of two
parameters --- a multiplier $S_\Sigma$ to transform the model column density to
the observed one, and a multiplier $S_r$ to rescale the linear size of the model.
Other rescaling coefficients can be derived from the above two parameters:
$S_\rho=S_\Sigma/S_r$ for density, $S_m=S_\Sigma S_r^2$ for mass, $S_v=(S_\Sigma
S_r)^{1/2}$ for velocity, $S_T=S_\Sigma S_r$ for initial gas temperature, etc.
For example, we can derive the scaling mass $M_s$ and scaling radius $r_s$ for
the DM halo.

The minimum $\chi^2$ values are much larger than
unity for all four galaxies (this is also true for best-fitting King models; see
Table~\ref{tab1}). The reasons for that are that (1) the observed surface
brightness profiles show many fine details which cannot be reproduced by any
simple model, and/or (2) the quoted measurement errors are underestimated.  As a
result, our strategy is to look for models producing smaller $\chi^2$ values
than the best fitting King models: $\chi'^2\equiv \chi^2/\chi^2_{\rm
King}\lesssim 1$.

\subsection{Plausibility of DM halos}
\label{par3}

To quantify the third criterion in \S\ref{par0}, we used the formula of
\citet{she99} to calculate the comoving number density of DM halos as a function of
virial mass $M_{\rm vir}$ and redshift $z$:

\begin{equation}
\label{ST}
\frac{dn}{d\ln M_{\rm vir}}= \frac{A}{\sqrt{2\pi}} \left(1+\frac{1}{\nu^{2q}}\right) 
\frac{\rho_m \nu}{S}\left|\frac{dS}{dM_{\rm vir}}\right| \exp\left(-\frac{\nu^2}{2}\right).
\end{equation}

\noindent Here $\nu\equiv (a/S)^{1/2} \delta(z)$, the
present day mass density of the universe is $\rho_m=3\Omega_m H^2/(8\pi G)$, and
the values of numerical coefficients are $A=0.322$, $a=0.707$, $q=0.3$;
$\Omega_m$ and $H$ are mass density of the universe in critical density units
and Hubble constant. We calculate the variance $S$ of the primordial density
field on mass scale $M_{\rm vir}$, extrapolated linearly to $z=0$, from fitting
formulae given in Appendix~A of
\citet{vdb02}, which are accurate to better than 0.5\% over the mass range
$10^6<M_{\rm vir}<10^{16}$~M$_\odot$. The shape parameter $\Gamma$, required for
the calculation of $S$, is estimated as 

\begin{equation}
\Gamma=\Omega_m h\exp[-\Omega_b(1+\sqrt{2h}/\Omega_m)]
\end{equation}

\noindent \citep{sug95}, where $h\equiv H/(100$~km~s$^{-1}$~Mpc$^{-1})$ and 
$\Omega_b$ is the baryonic mass density of the universe in critical density
units.  The critical overdensity threshold for spherical collapse, linearly
extrapolated to $z=0$, is approximated by

\begin{equation}
\delta(z)=0.15(12\pi)^{2/3}\Omega_m^{0.0055} / D(z)
\end{equation}

\noindent \citep{NFW97}, which is valid for flat $\Lambda$CDM cosmologies.
The fitting formula for the linear growth factor $D(z)$ was taken from
Appendix of \citet{NFW97}. Throughout this paper, we adopt the following values of
cosmological constants: $\Omega_m=0.27$, $\Omega_b=0.044$,
$H=71$~km~s$^{-1}$~Mpc$^{-1}$, and $\sigma_8=0.84$
\citep{spe03}. We assume a flat $\Lambda$CDM cosmology: $\Omega_\Lambda=1-\Omega_m$.

Fitting model surface brightness profiles to the observed ones gives us the
values of the scaling halo mass $M_s$ and scaling halo radius $r_s$ (see
\S~\ref{par1}). To use equation~(\ref{ST}), we need to know $M_{\rm vir}$ and
$z$. To make a transformation from ($M_s$,$r_s$) to ($M_{\rm vir}$,$z$), we use
the fact that in cosmological $\Lambda$CDM simulations the halo concentration
$c=r_{\rm vir}/r_s$ distribution is approximately lognormal (for fixed $M_{\rm
vir}$ and $z$), with dispersion $0.14$~dex
\citep{bul01}. For low mass halos (with $M_{\rm vir}=10^8\dots 10^{11}$~M$_\odot$),
halo concentration depends on $M_{\rm vir}$ and $z$ in the following way
\citep{bul01,ste02}:

\begin{equation}
\label{eqc}
c=\frac{27}{1+z} 10^{0.14\nu_c} \left(\frac{M_{\rm vir}}{10^9 M_\odot}\right)^{-0.08}.
\end{equation}

\noindent Here $\nu_c$ is the number of standard deviations from the mean 
concentration. Multiplying 
$dn/d\ln M_{\rm vir}$ (eq.~[\ref{ST}]) by the probability to find a halo
with $\nu_c'=\nu_c\dots\nu_c+d\nu_c$, $(2\pi)^{-1/2} \exp(-\nu_c^2/2)d\nu_c$, and dividing
the result by $d\nu_c$ gives us the comoving halo number density per unit
$\ln M_{\rm vir}$ and per standard deviation in concentration:

\begin{eqnarray}
\label{ST2}
F\equiv\frac{dn}{d\ln M_{\rm vir} d\nu_c}= \frac{A}{2\pi} \left(1+\frac{1}{\nu^{2q}}\right) 
\frac{\rho_m \nu}{S}\left|\frac{dS}{dM_{\rm vir}}\right|\times&&\nonumber\\
\times \exp\left(-\frac{\nu^2+\nu_c^2}{2}\right).&&
\end{eqnarray}

We use equation~(\ref{ST2}) to find the most probable combination of $M_{\rm vir}$ and $z$
for a halo with given $M_s$ and $r_s$. Our algorithm is as follows. We vary $\nu_c$ in
the interval $-4\dots 4$ with a small step (0.01). For each value of $\nu_c$, the following
non-linear equation is solved numerically to find $z$:

\begin{equation}
\label{zeq}
M_{\rm vir}(z)=M_s M'\{c[M_{\rm vir}(z),z]\}.
\end{equation}

\noindent Here 

\begin{eqnarray}
\label{zeq2}
M_{\rm vir}(z)=\left[(27 r_s 10^{0.14\nu_c})^3 (10^9 {\rm M}_\odot)^{0.24}\times\right.&&\nonumber\\
\left.\times\frac{\Delta_c H^2 \Omega_m}{2G \Omega_{m,z}}\right]^{1/(1+0.24)}&&
\end{eqnarray}

\noindent was derived from the definition of critical overdensity for
collapsed halos

\begin{equation}
\label{zeq3}
\Delta_c=\frac{2GM_{\rm vir}\Omega_{m,z}}{r_{\rm vir}^3 H^2 \Omega_m (1+z)^3}
\end{equation}

\noindent and equation~(\ref{eqc}) for halo concentration 
$c$; $M'(x)$ is given by equation~(\ref{Ms}). The mass density in flat universe
as a function of redshift is $\Omega_{m,z}=\{1+\Omega_\Lambda/[\Omega_m
(1+z)^3]\}^{-1}$. We use the fitting formula $\Delta_c\simeq 18\pi^2+82d-39d^2$
from \citet{bry98}, which is accurate to 1\% in the range
$\Omega_{m,z}=0.1-1$. (Here $d\equiv \Omega_{m,z}-1$.)  \

After solving equation~(\ref{zeq}) (which gives us $z$ for given $M_s$, $r_s$,
and $\nu_c$), we use equation~(\ref{zeq2}) to find the corresponding $M_{\rm
vir}$ value. Finally, from equation~(\ref{ST2}) we estimate the comoving halo
number density $F(\nu_c)$. Repeating the above procedure for the whole interval
of $\nu_c$, we find the $\nu_c$ value (and corresponding values for $z$ and $M_{\rm
vir}$) which maximizes the likelihood function $F(\nu_c)$.

We make the following very rough estimate of the lower cutoff $F_{\rm min}$
value (such as that the halos with $F\ll F_{\rm min}$ would be very unlikely
progenitors of dSphs, whereas the halos with $F\gtrsim F_{\rm min}$ would be
likely progenitors). We divide the estimated mass of the Local Group of $\sim
2.3\times 10^{12}$~M$_\odot$ \citep{ber99} by the present day mass density of
the universe $\rho_m$ to derive the comoving volume of the part of the early
universe which became the Local Group, $V_{\rm LG}\sim 61$~Mpc$^3$.  (This volume
corresponds to a sphere with comoving radius of 2.4~Mpc.)  There are $N_{\rm
dSph}\sim 20$ dSph galaxies in the Local Group. From the analysis presented in
this paper, the DM halo masses for these objects span $\sim 2$~dex, so $\Delta
\ln M_{\rm vir}\sim 4.6$. When the very first of these halos was virialized, the
corresponding comoving number density was $F\sim 1/(V_{\rm LG} \Delta \ln M_{\rm
vir})$. Much later on, when all $N_{\rm dSph}$ halos had formed, the comoving
number density became $F\sim N_{\rm dSph}/(V_{\rm LG} \Delta \ln M_{\rm vir})$
(which would only be accurate if the halos did not accrete any mass after their
formation). To come up with a single number, we take an average (in log) of the
two above extremes: $\langle F\rangle\sim N^{1/2}_{\rm dSph}/(V_{\rm LG} \Delta
\ln M_{\rm vir})\sim 0.016$~Mpc$^{-3}$.  We adopt $F_{\rm
min}=0.01$~Mpc$^{-3}$ for the rest of this paper.

\subsection{Strategy of Searching for Best Fitting Models}

\begin{figure*}
\epsscale{.80}
\plottwo{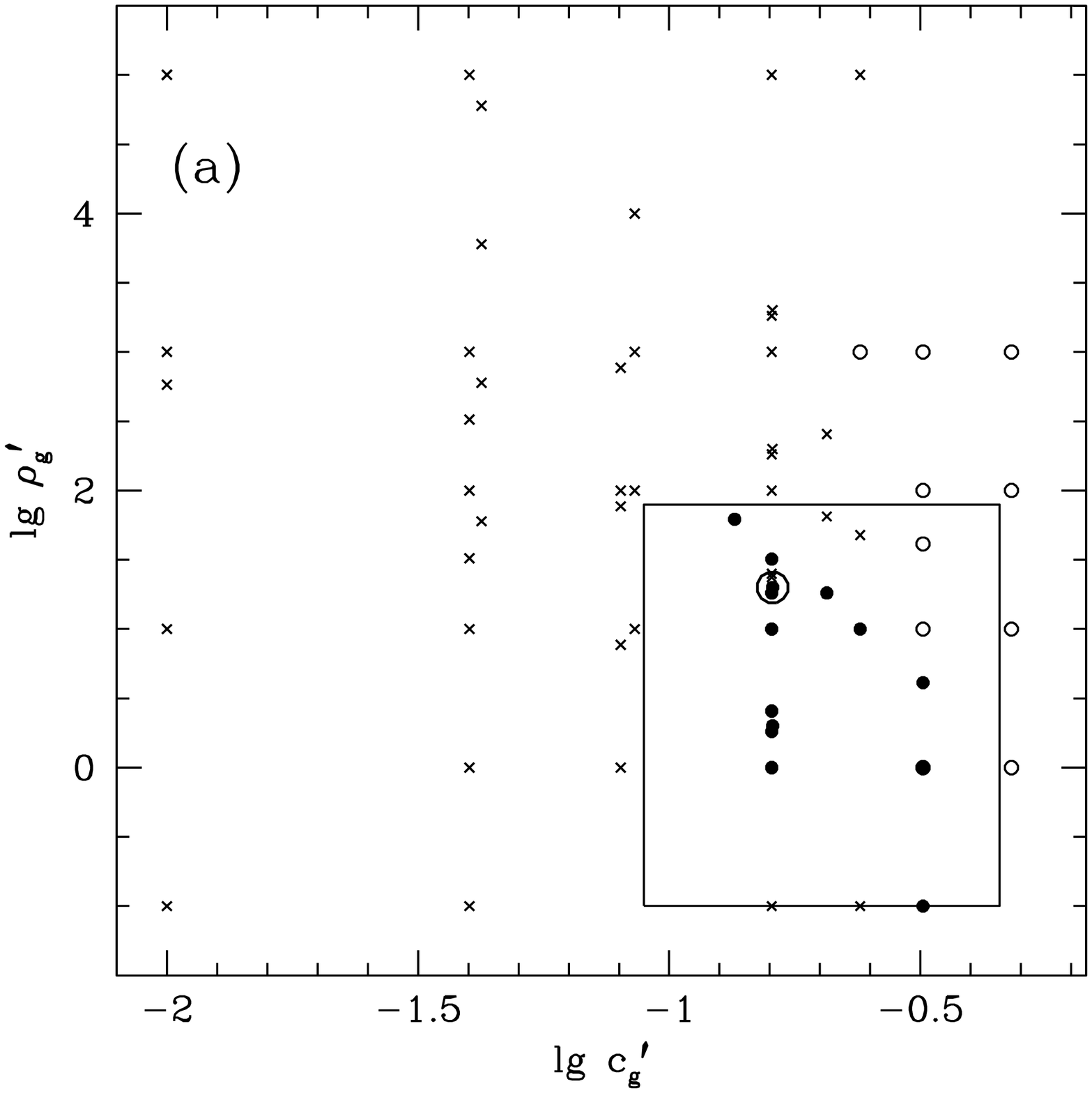}{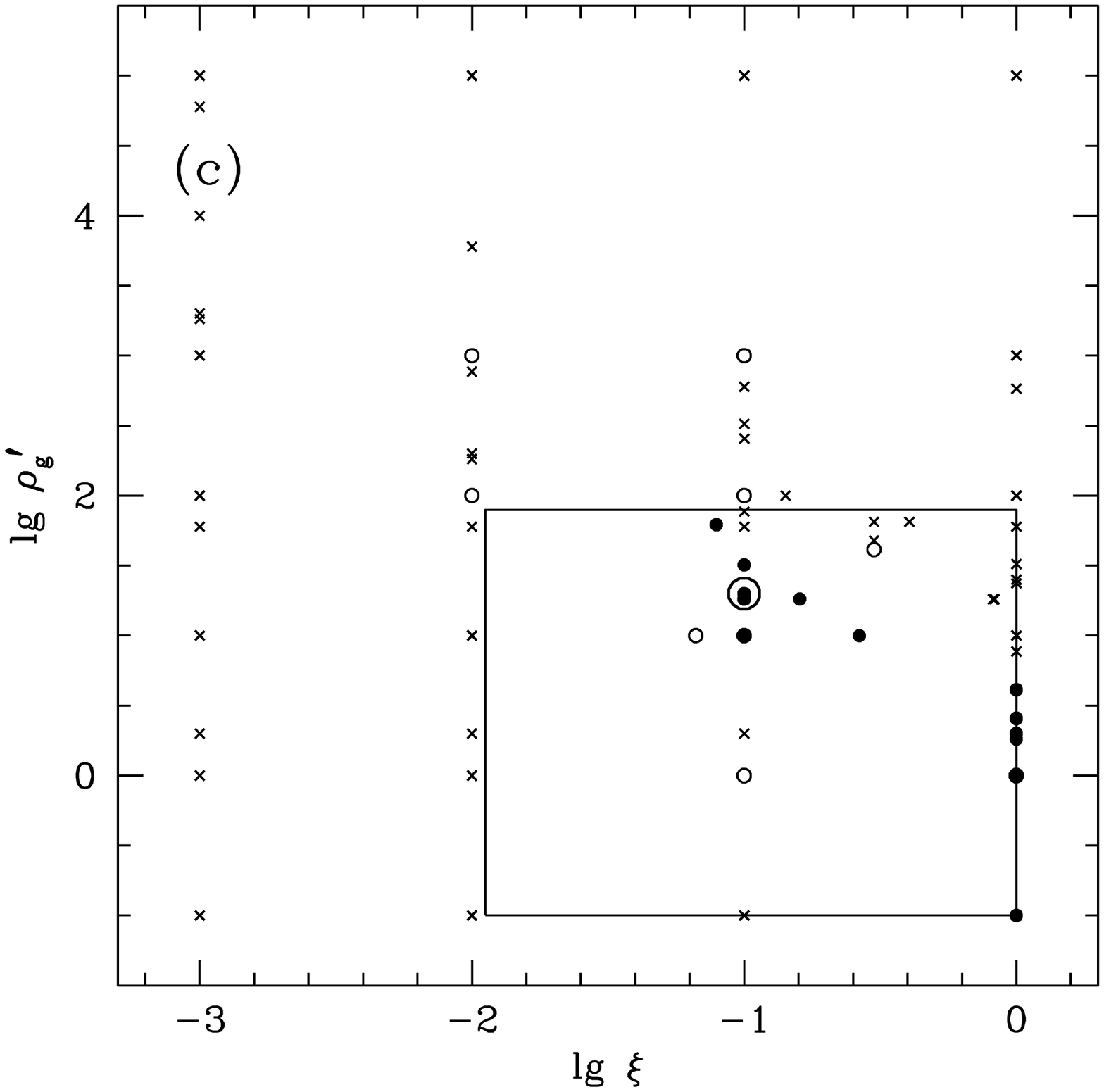}
\plottwo{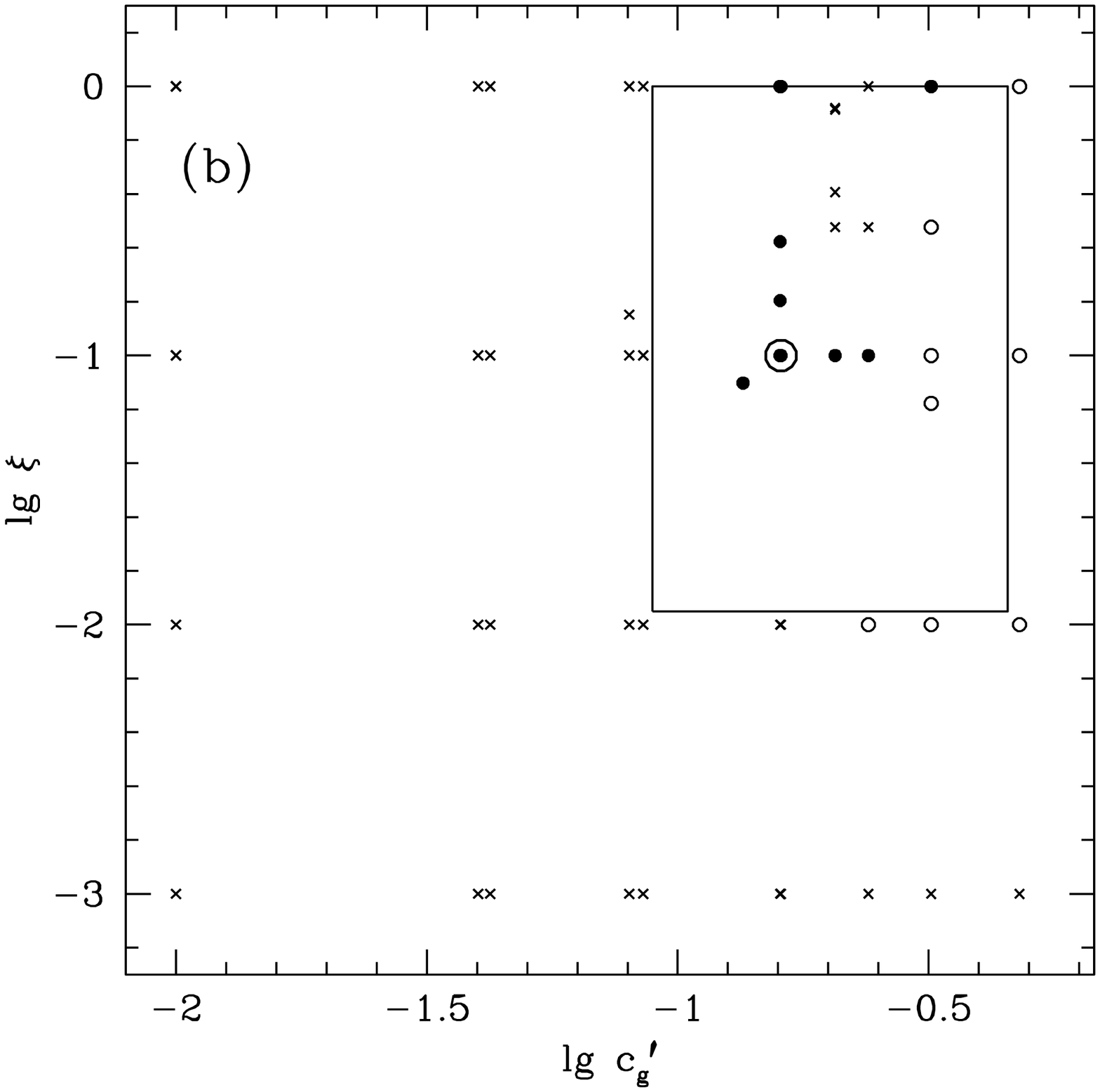}{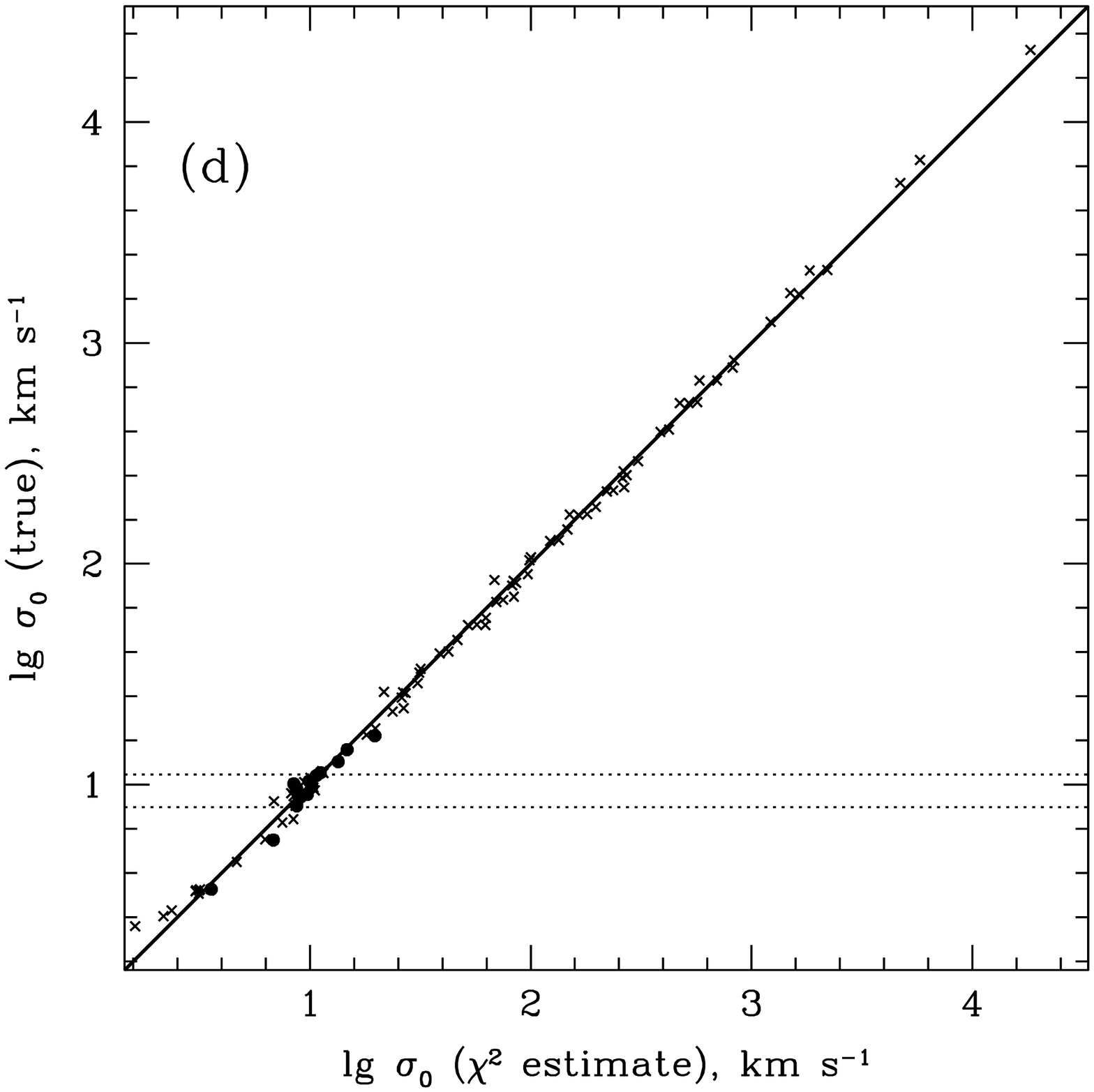}
\caption {Illustration of our best fitting model search algorithm for the Draco1 dataset
(see Section~\ref{results}), NFW halos and $\lambda=0.1$. Panels (a-c) show
the projection of the models onto three orthogonal planes in the three-dimensional
space of free model parameters $c_g'$, $\rho_{g,0}'$, and $\xi$. Crosses mark
models which have unphysical DM halos ($F<F_{\rm min}$).  Small open circles
correspond to physical ($F>F_{\rm min}$) models which have a bad $\chi^2$ fit
between the modeled and observed surface brightness profiles ($\chi^2>2\chi_{\rm
min}^2$, where $\chi_{\rm min}^2$ is the lowest $\chi^2$ value for all $\sim
700$ models). Solid circles show the location of models which both are physical
($F>F_{\rm min}$) and fit well the observed surface brightness profile
($\chi^2<2\chi_{\rm min}^2$). The large circle marks a physical model with
the smallest $\chi^2$ value (for NFW halo and $\lambda=0.1$). Rectangular boxes
delineate the area where good models are being looked for.  Panel (d) shows the
multivariate correlation between the three free parameters ($c_g'$,
$\rho_{g,0}'$, and $\xi$) and the core line-of sight velocity dispersion
$\sigma_0$ for models with $\chi^2<2\chi_{\rm min}^2$, which is used for
searching for good models.  Here crosses correspond to unphysical DM halos (with
$F<F_{\rm min}$), whereas the solid circles mark physical halos ($F>F_{\rm
min}$). Two horizontal dotted lines show the one-sigma errorbars for the
observed $\sigma_{\rm obs,0}$ in Draco.
\label{fig3} }
\end{figure*}

In this project we faced the formidable task of sampling 4-dimensional ($c_g'$,
$\rho_{g,0}'$, $\xi$, and $\lambda$) initial parameter space in our numerical
simulations (with an extra factor of 2 due to the two different types of DM halo
profiles considered --- NFW and Burkert). As we mentioned before, the
requirement for gas to be colder than the virial temperature results in
$0<c_g'\lesssim 0.5$.  Other obvious constraints are $0<\xi\leqslant 1$
and $0 \leqslant\lambda < 1$ (by definition). We had no other reasonably
motivated constraints on the range of our free parameters.

Initially, we sampled a large range of the free parameter space by simulating a
grid of models with $c_g'=(0.01,0.04,0.16)$, $\rho_{g,0}'=(0.1,10,10^3,10^5)$,
$\xi=(0.001,0.01,0.1,1)$, and $\lambda=(0,0.1)$, which resulted in 192 models
(both NFW and Burkert). We quickly realized that models with $c_g'<0.04$ or
$\rho_{g,0}'>10^3$ can be safely ruled out as they all produced unacceptably
large $\chi^2$ values (in other words --- their surface brightness profiles were
nothing like profiles of dSphs) and/or very low $F$ values being much smaller
than $F_{\rm min}$ (meaning that the inferred DM halo parameters were improbable
in the adopted $\Lambda$CDM cosmology) for our sample of galaxies.

Other important observations were: (1) the range for $c_g'$ should be expanded
up to $\sim 0.5$ to include warmer models (which we could not do for models with
$\lambda=0$, as in most of such models the initial stellar half-mass radius
becomes divergent for $c_g'\gtrsim 0.2$); (2) in the range $0.04\dots 0.5$, the
parameter $c_g'$ should be better sampled (by a factor of two) ; (3) in the
range $0.1\dots 10^3$, $\rho_{g,0}'$ should also be better sampled (by a factor
of two).

As a result, we ran 294 additional models, which also included a set of models
with $\lambda=0.3$ (for such models we simulated only the cases with
$c_g'\geqslant 0.04$ and $\rho_{g,0}'\leqslant 10^3$). Altogether, we had 486
models on a regularly spaced 4-dimensional grid.

The problem we faced at that point was that no realistic number of models on a
regularly sampled 4-dimensional grid would be large enough to simultaneously
satisfy all our three selection criteria (outlined in \S~\ref{par0}). More
specifically, despite the fact that for most of our galaxies a relatively large
volume in the 4-dimensional parameter space would satisfy both the first and the
third criteria from \S~\ref{par0}, few (usually none) of the models from this
volume would produce the core line-of-sight velocity dispersion $\sigma_0$ which
would be consistent with observations within the measurement errors (our second
criterion).

Fortunately, we found a solution for the above problem. We noticed that for the
models which produce reasonably good $\chi^2$ fits (within a factor of two from
the model with the lowest $\chi^2$ value), there is a good
multivariate correlation (in logarithmic space) between $\sigma_0$ and the free
parameters $c_g'$, $\rho_{g,0}'$, and $\xi$ for NFW models, and $\rho_{g,0}'$
and $\xi$ for Burkert models (the parameter $\lambda$ is fixed).

Our strategy was then as follows. For each galaxy, each value of $\lambda$, and
each type of DM profile (NFW or Burkert), we used a multivariate $\chi^2$
fitting to find coefficients for the $\sigma_0(c_g',\rho_{g,0}',\xi)$
correlation. (Only the models producing good surface brightness $\chi^2$ fits
were used.) We used this correlation to randomly sample a plane
$\sigma_{{\rm obs},0}=\sigma_0(c_g',\rho_{g,0}',\xi)$, with typically 10
additional models for each case. (Here $\sigma_{{\rm obs},0}$ is the observed
core line-of-sight velocity dispersion, listed in Table~\ref{tab1}.) The
sampling was done within the three-dimensional rectangular volume corresponding
to models which simultaneously satisfied our first and third criteria from
\S~\ref{par0}. As our method is not precise, not all additional models satisfied 
all the three criteria. Figure~\ref{fig3} illustrates our algorithm for the case
of the Draco1 dataset (see Section~\ref{results}), NFW halos, and the $\lambda=0.1$.

Overall, we simulated around 700 models. It took approximately 500 CPU-days to
simulate all the models on McKenzie cluster at CITA composed of 256 dual Intel
Xeon 2.4 GHz nodes networked with commodity gigabit ethernet. Each model was
run on 4 CPUs, with many models running in parallel.

\section{RESULTS FOR FOUR DWARF SPHEROIDAL GALAXIES}
\label{results}

\subsection{General Remarks}

In Table~\ref{tab2} we show statistics on ``good'' models (which meet all three
selection criteria from \S~\ref{par0}) for five datasets considered in this paper
--- for the Draco (two datasets), Sculptor, Carina, and Fornax dSphs. Because of the
small number of ``good'' models $N_g$, the not completely random nature of our best
fitting models search algorithm, and some degree of noisiness in the results of our
simulations, in this table we present three complimentary statistical measures for
each parameter and for each dataset: the value for the ``best'' model (producing
the smallest value of $\chi^2$ in surface brightness profile fitting), the mean
value with the associated standard deviation, and the total range for all $N_g$
models. Most of the parameters are presented as their logarithms as their
distribution is much less skewed in logarithmic
space. For the same reason, instead of the redshift, $z$, of halo formation, we present
statistics on the cosmic scale factor $a=1/(1+z)$.

We use the following definition of tidal radius $r_{\rm tid}$ \citep{hay03}:

\begin{equation}
\frac{M(r_{\rm tid})}{r_{\rm tid}^3} = \left[ 2-\frac{R}{M_H(R)}\frac{\partial M_H}{\partial R}\right]
\frac{M_H(R)}{R^3}.
\label{eq_rtid}
\end{equation}

\noindent Here $M(r)$ and $M_H(R)$ are enclosed mass functions for the satellite 
and host galaxies. We assumed that Milky Way DM halo has NFW density profile
with $M_{\rm vir}=1.5\times 10^{12}$~M$_\odot$ (corresponding to $R_{\rm
vir}=298$~kpc). From equation~(14) of \citet{ste02}, the concentration of the
Milky Way halo is $c=13.2$. The tidal radius was calculated for the current
distances $D$ (listed in Table~\ref{tab1}) of the dSphs from the Galactic
center.

To calculate the temperature $T$ of the gas, we assumed that the mean molecular
weight is $\mu=1.218 m_p$, which is appropriate for neutral gas of primordial
composition. (Here $m_p$ is the mass of the proton.) If gas is ionized, the
temperature would be a factor of 2 smaller.

In the next four sections we also present surface density $\Sigma(r)$ and
line-of-sight velocity dispersion $\sigma(r)$ profiles for our best fitting
models for each dataset (Figures~\ref{fig4}-\ref{fig8}). Only those parts of the
profiles which have a standard deviation (estimated from the last $N_S$ snapshots for a given
radial bin) smaller than 0.2~dex for $\Sigma(r)$ and 1~km~s$^{-1}$ for
$\sigma(r)$ are shown.  For comparison, we show the surface density profiles for
the best fitting theoretical \citet{K66} models (dashed lines).

\subsection{Draco}
\label{Dra}

\begin{figure*}
\plottwo{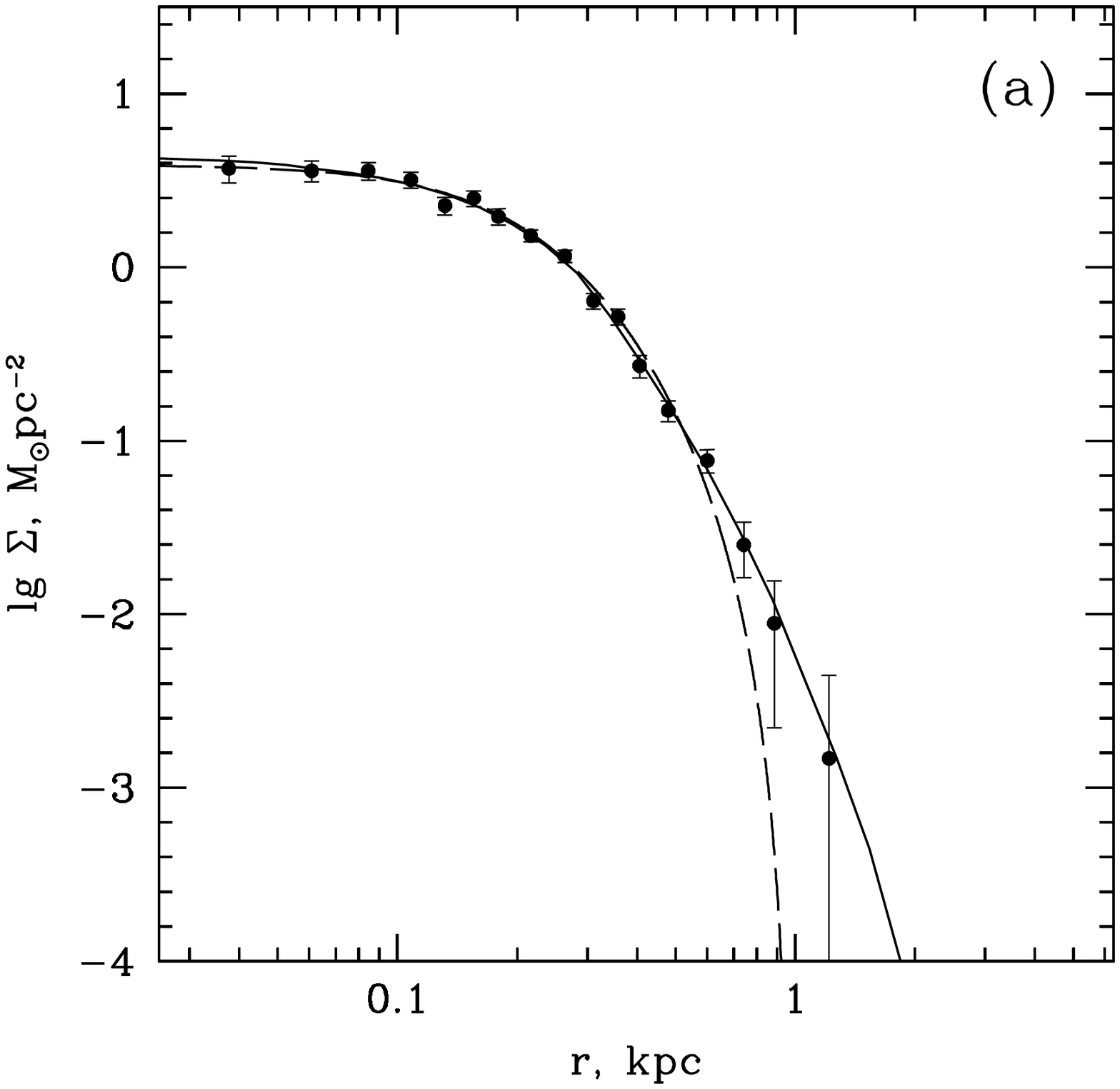}{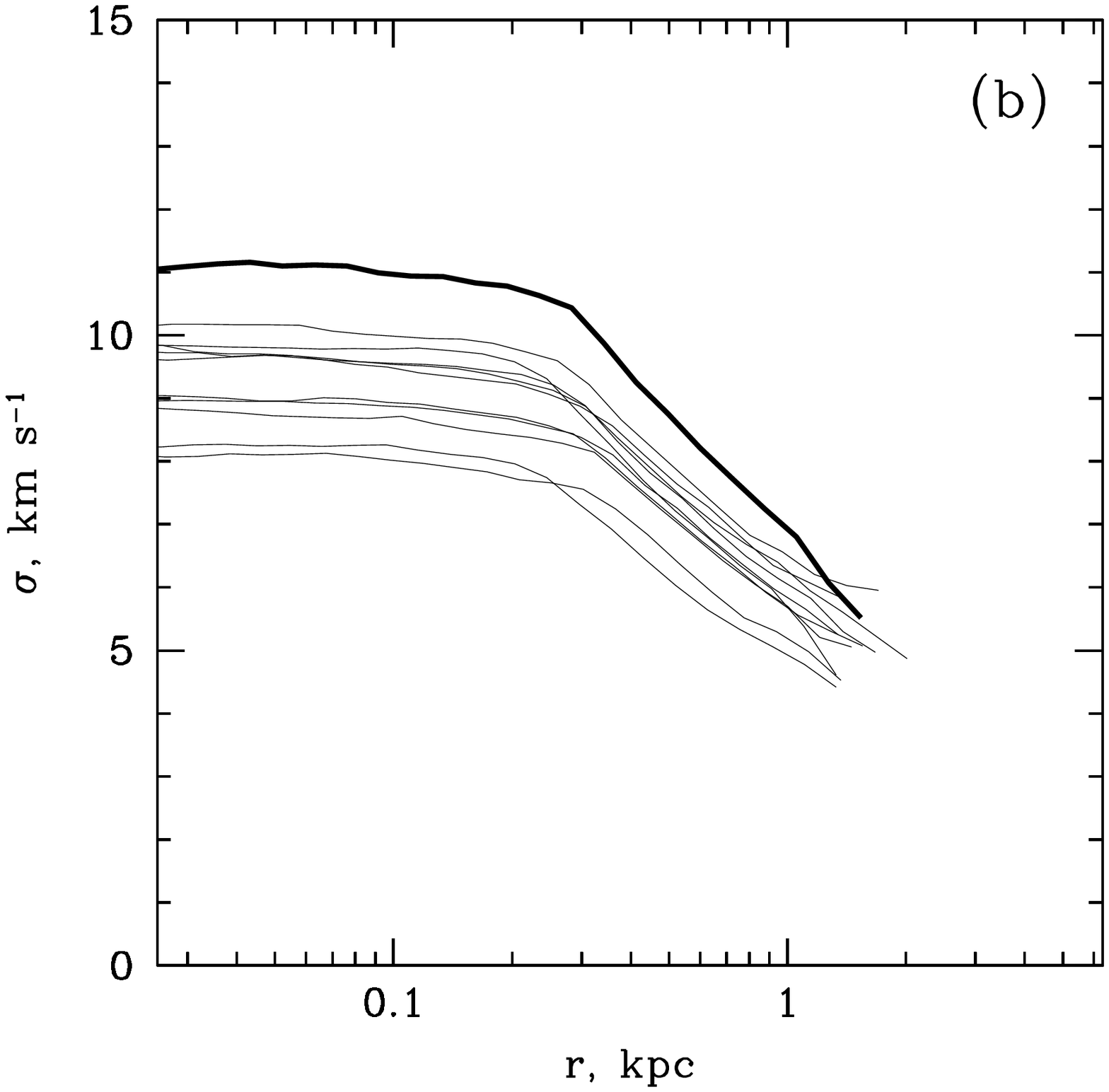}
\caption {Best fitting models for the Draco1 dataset. (a) Surface density profiles. The observed profile is shown
as solid circles with errorbars. The solid line corresponds to the best fitting
model (with the lowest $\chi'^2$).  The dashed line shows the best fitting
theoretical King model profile. (b) Line-of-sight velocity dispersion profiles
for all 11 ``good'' models from Table~\ref{tab2}. The thick line corresponds to
the model with the lowest $\chi'^2$.
\label{fig4} }
\end{figure*}

\begin{figure*}
\plottwo{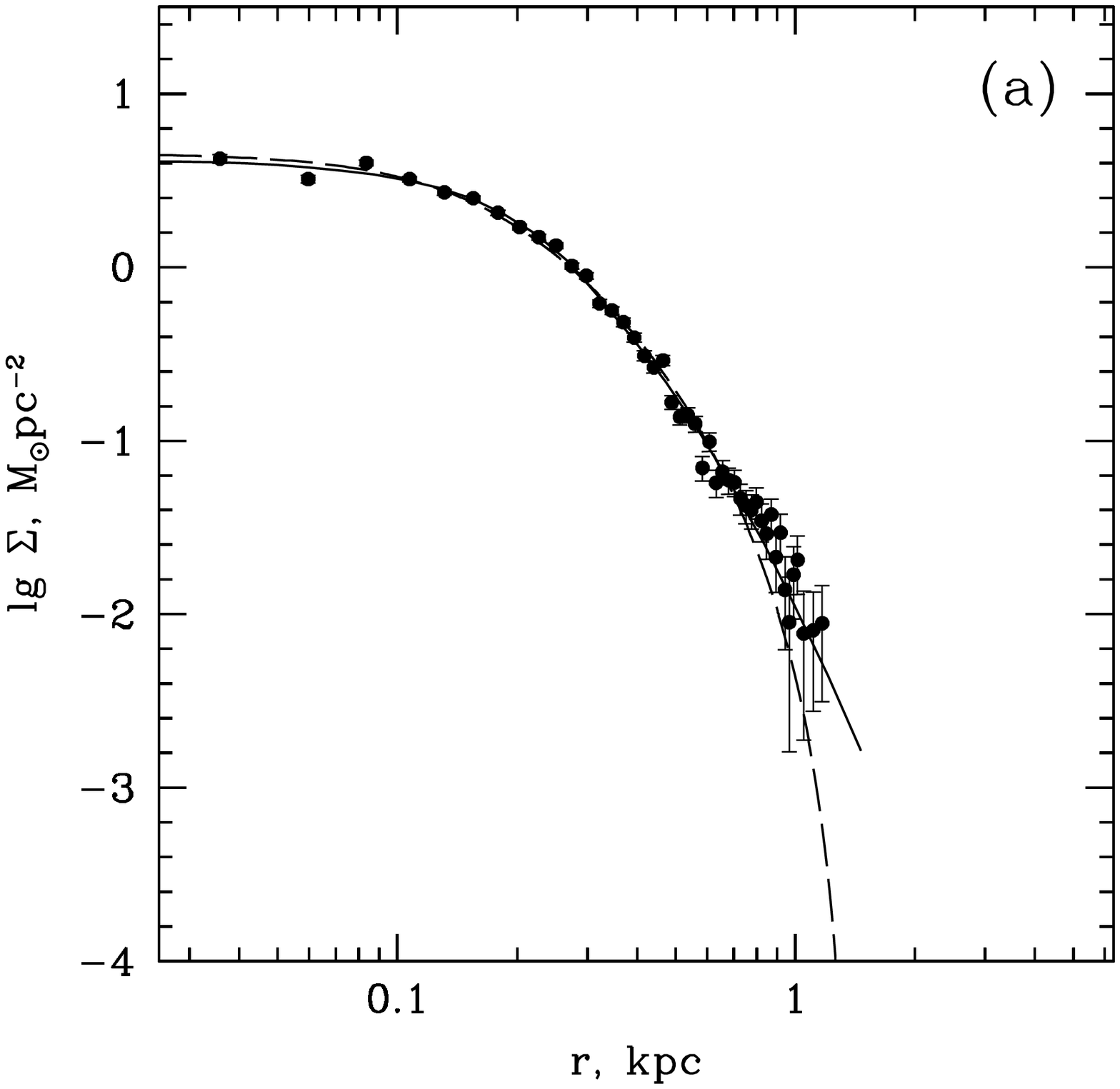}{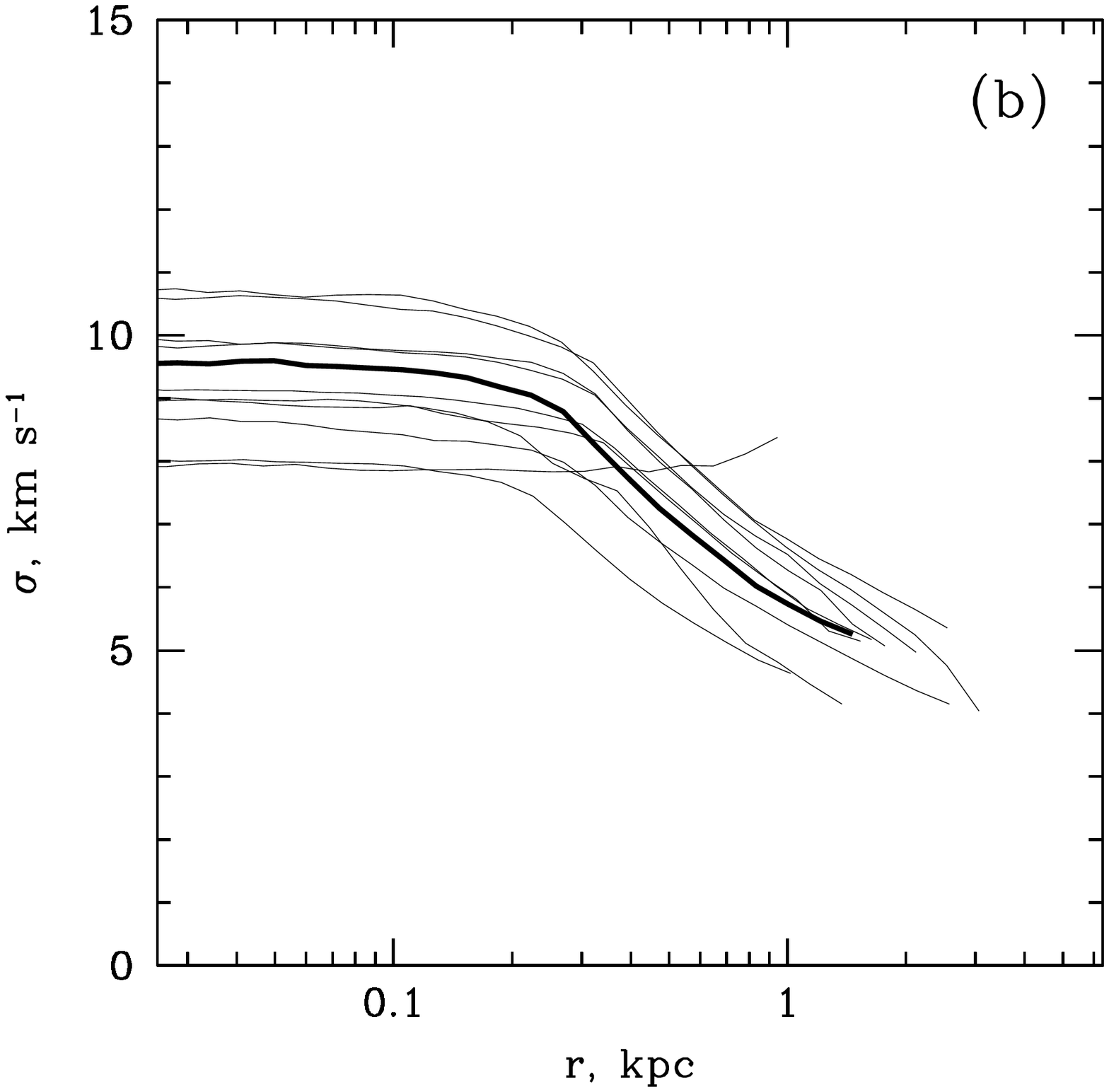}
\caption {Best fitting models for the Draco2 dataset. (a) Surface density profiles. The observed profile is shown
as solid circles with errorbars. The solid line corresponds to the best fitting
model (with the lowest $\chi'^2$).  The dashed line shows the best fitting
theoretical King model profile. (b) Line-of-sight velocity dispersion profiles
for all 13 ``good'' models from Table~\ref{tab2}. The thick line corresponds to
the model with the lowest $\chi'^2$.
\label{fig5} }
\end{figure*}

Draco is probably the best candidate for having an extended DM halo among Galactic
dSph satellites. \citet{ode01} used deep multicolor photometry covering 27
square degrees around the center of this galaxy to show that Draco has a very
regular structure (with no signs of tidal interaction with the Milky Way
gravitational field) out to the radius where its surface brightness falls to 0.003 of
the central value. Their results suggested that there is much more DM in the
outskirts of Draco than in its center.

More recently, \citet{wil04} presented a new star number count profile for
Draco.  The new profile appears to be even more shallow in the outskirts of the
galaxy than the profile of \citet{ode01}, though the two profiles are marginally
consistent. \citet{wil04} also published a line-of-sight velocity dispersion
profile for Draco, which is roughly isothermal, with the exception of the last
radial bin where the stars are marginally colder than the rest of the galaxy.

We use star count profiles from both \citet[their sample S2, which has a lower
estimated foreground/background contamination level than their sample S1]{ode01} and
\citet{wil04} in our analysis, which we designate Draco1 and Draco2,
respectively. We found 11 and 13, respectively, ``good'' models for these two datasets (see
Table~\ref{tab2}). In both cases, our models produce much better $\chi^2$
surface brightness fits than best fitting King models ($\chi'^2\equiv
\chi^2/\chi^2_{\rm King}\sim 0.6$). As can be seen in Figures~\ref{fig4}a and
\ref{fig5}a, the only statistically significant difference between
our best fitting models and King models is in the outer parts of the galaxy.  In
the Draco outskirts, the King models exhibit a sharp cutoff due to the external
tidal field, whereas our models have a relatively shallow power-law profile,
which is completely consistent with the observed profiles. Virtually all of our
best fitting models have non-zero value for the star formation lower density
cutoff $\lambda$, which results in the outskirts of the galaxy appearing to be
dynamically colder in the model $\sigma(r)$ profile, consistent with
the observations of \citet{wil04}. As we showed in \S~\ref{props}, the total
velocity dispersion in our models with $\lambda>0$ does not decrease with
radius, with the observed decline of the line-of-sight velocity dispersion being
due to significant radial anisotropy exhibited by our models outside the
initial (non-equilibrium) stellar radius $x_\lambda$ (see Figure~\ref{fig2}b).

As can be seen from Table~\ref{tab2}, the best fitting models for Draco have a large
range of DM halo virial masses $\sim 10^8- 10^{10}$~M$_\odot$. These masses are all
significantly larger than conventional estimates of $\sim 2\times 10^7$~M$_\odot$
\citep{mat98} which are based on the assumption that ``mass follows light''.
The total range for halo formation redshifts is $\sim 1.3-10$, corresponding to
a galaxy age $\sim 9-13$~Gyr, which is consistent with the estimated age of
the oldest Draco stars of $10-14$~Gyr \citep{mat98}.

It is interesting to note that a comparable number of both NFW and Burkert
models provided good fits to the observed Draco profiles. (Out of 24 ``good''
models for Draco1 and Draco2 datasets, there are 16 NFW and 8 Burkert profiles.)
Even more interesting is the fact that for 16 ``good'' NFW models $\nu_c\simeq
-1.21\pm 0.44$, whereas for 8 Burkert models $\nu_c\simeq 1.16\pm 0.31$. It
appears that the Draco galaxy is best described by models with either underdense NFW
or overdense Burkert DM halos, which is consistent with
Draco having a DM halo with an intermediate value of its central density slope
of $\gamma\sim -0.5$.

The predictions of our models for the initial gas state in Draco appear to be
sensible, with the temperature of $\sim 20,000$~K ($\sim 10,000$~K if the gas
was ionized) and the central number density of $\sim 10$~cm$^{-3}$. The
lowest density for star-forming gas is $\sim 1-2$~cm$^{-3}$. After the
starburst, the stellar cluster in all ``good'' models expanded by a very modest factor
of $\sigma_{\rm ini,0}/\sigma_0\sim 1.2$. 

Overall, our simple model of single-burst star formation from isothermal gas in
an extended DM halo appears to provide an adequate description of the Draco dSph.

\subsection{Sculptor}

\begin{figure*}
\plottwo{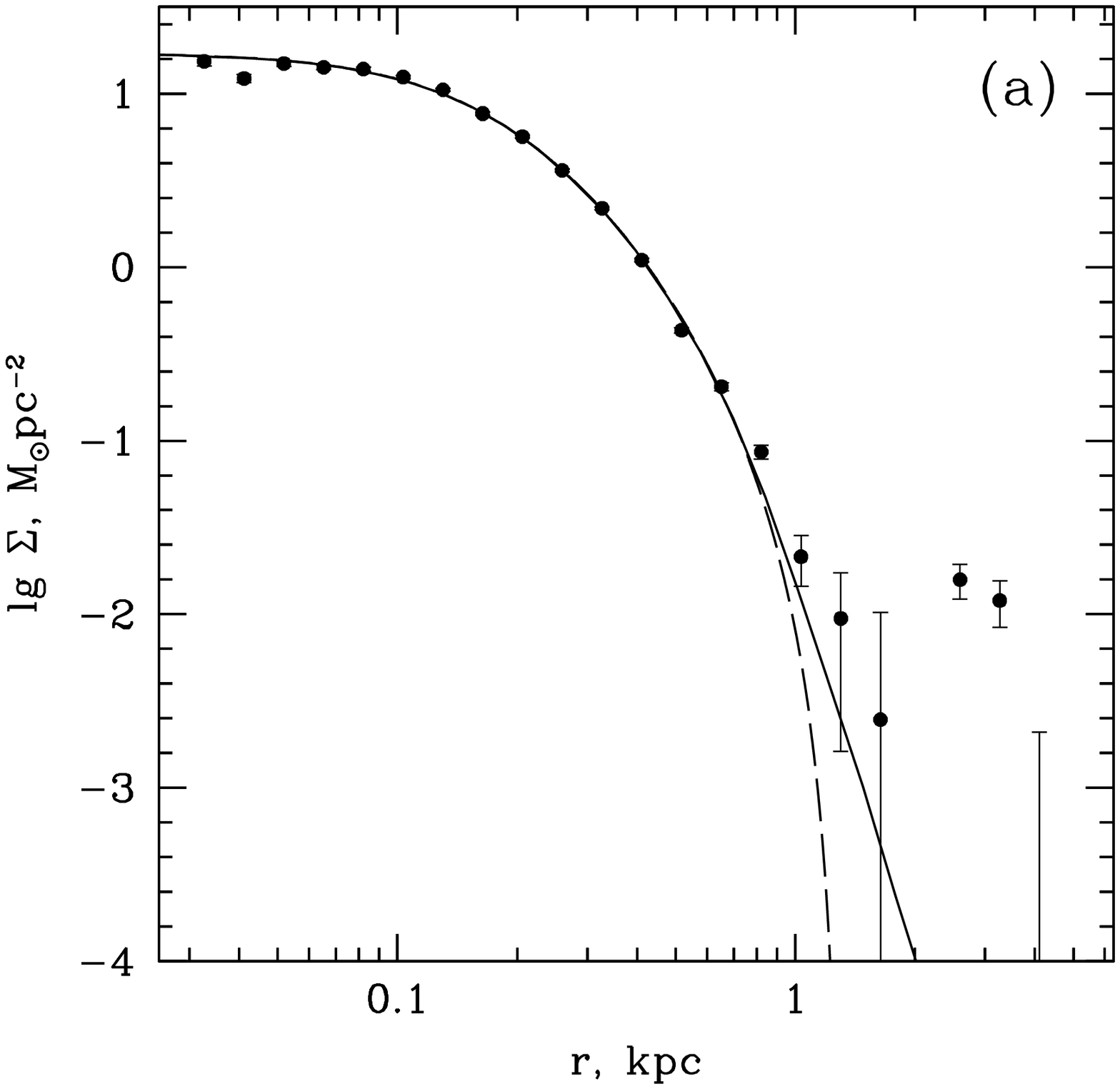}{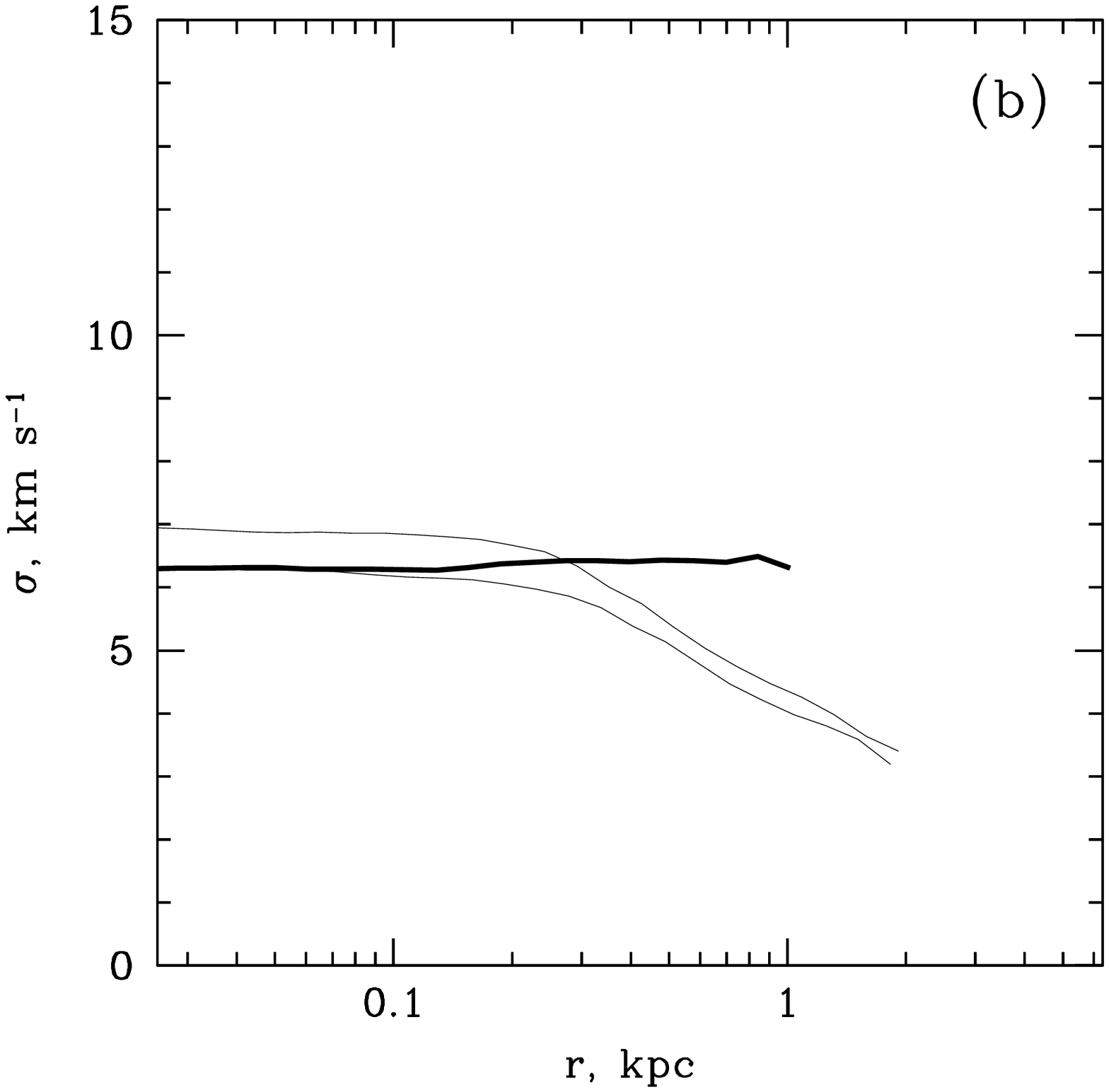}
\caption {Best fitting models for Sculptor. (a) Surface density profiles. The observed profile is shown
as solid circles with errorbars. The solid line corresponds to the best fitting
model (with the lowest $\chi'^2$).  The dashed line shows the best fitting
theoretical King model profile. (b) Line-of-sight velocity dispersion profiles
for all 3 ``good'' models from Table~\ref{tab2}. The thick line corresponds to
the model with the lowest $\chi'^2$.
\label{fig6} }
\end{figure*}

For Sculptor (and also for Carina and Fornax) we used star count profiles from
\citet{wal03}. The authors use single-color photometry to identify stars belonging
to these galaxies. As a result, the star count profiles for these three dSphs
are of much lower quality than the Draco data of \citet{ode01}, who employed
a multicolor mask filtering technique to improve the sensitivity to Draco stars
by a factor of $\sim 30$. One has to keep in mind that simple single-color
techniques [such as used by \citet{wal03}] can produce very unreliable results
at surface brightness levels below the foreground/background contamination
level. For example, according to the single-color photometry of
\citet{irw95}, Draco has a flat surface brightness profile beyond 20~arcmin from
the galactic center (which was interpreted as a halo of unbound stars).  In
contrast, much higher sensitivity multicolor observations of \citet{ode01}
failed to find any evidence for such a halo.

In all the three galaxies (Sculptor, Carina, and Fornax) some of the radial bins
in their outskirts have negative surface brightness values (\citealt{wal03};
this is an artifact of background subtraction procedure). In the case of
Sculptor, even the upper one-sigma errorbar is negative for the bin at $r\sim
2$~kpc. As a result, we could not show this bin value in Figure~\ref{fig6}a. Our
$\chi^2$ fitting is not affected by the negative values of surface brightness,
as we perform the cross-correlation in linear space.

With all the above caveats in mind, our model appears to do a reasonably good
job in describing the properties of Sculptor. Formally, we have only one
``good'' model for this galaxy, with $\chi'^2=0.94$ (see Table~\ref{tab2}). On
the other hand, we have a large number of models with $\chi^2$ slightly worse
than for the best fitting King model: 3 models with $\chi'^2<1.1$, and 11 models
with $\chi'^2<1.5$. As will be seen in
\S~\ref{For}, this is very different from the case of Fornax, where the best
model has $\chi'^2\sim 1.1$, and the second-to-best has $\chi'^2\sim 1.9$.

Analysis of Figure~\ref{fig6}a shows that our best fitting model improves upon
theoretical King model by having a less steep outer surface brightness profile,
which is the same situation as with Draco (see \S~\ref{Dra}).  One cannot have
a much better fit with a smooth function due to the fact that the data are very
noisy in the outer parts of the galaxy, with the star counts of $-0.31\pm
0.02$~arcmin$^{-2}$ at $r\sim 2.1$~kpc and $0.09\pm 0.02$ in the next radial bin
at $r\sim 2.6$~kpc.

Properties of three models with $\chi'^2<1.1$ are summarized in
Table~\ref{tab2}. With such small statistics it is hard to draw any conclusions
on the range of model parameters, but it appears that all the main parameters
for Sculptor are comparable to those for Draco. Most importantly, the virial
mass for the DM halo is required to be $\gtrsim 3\times 10^8$~M$_\odot$, which is
much larger than the $6.4\times 10^6$~M$_\odot$ estimate of \citet{mat98} based on
the $\Upsilon=constant$ assumption. (Interestingly, the two \HI clouds observed in
the vicinity of Sculptor, which were argued by \citet{BCM03} to be a part of
ISM of this galaxy, are located in projection well within the inferred large
tidal radius of $\sim 7.5$~kpc, corresponding to $\sim 5.5$ angular degrees at
the distance $D=79$~kpc).  The notable difference from the Draco case is a
factor of two lower temperature, which is of order of 10,000~K for neutral gas
(see Table~\ref{tab2}).

\subsection{Carina}

\begin{figure*}
\plottwo{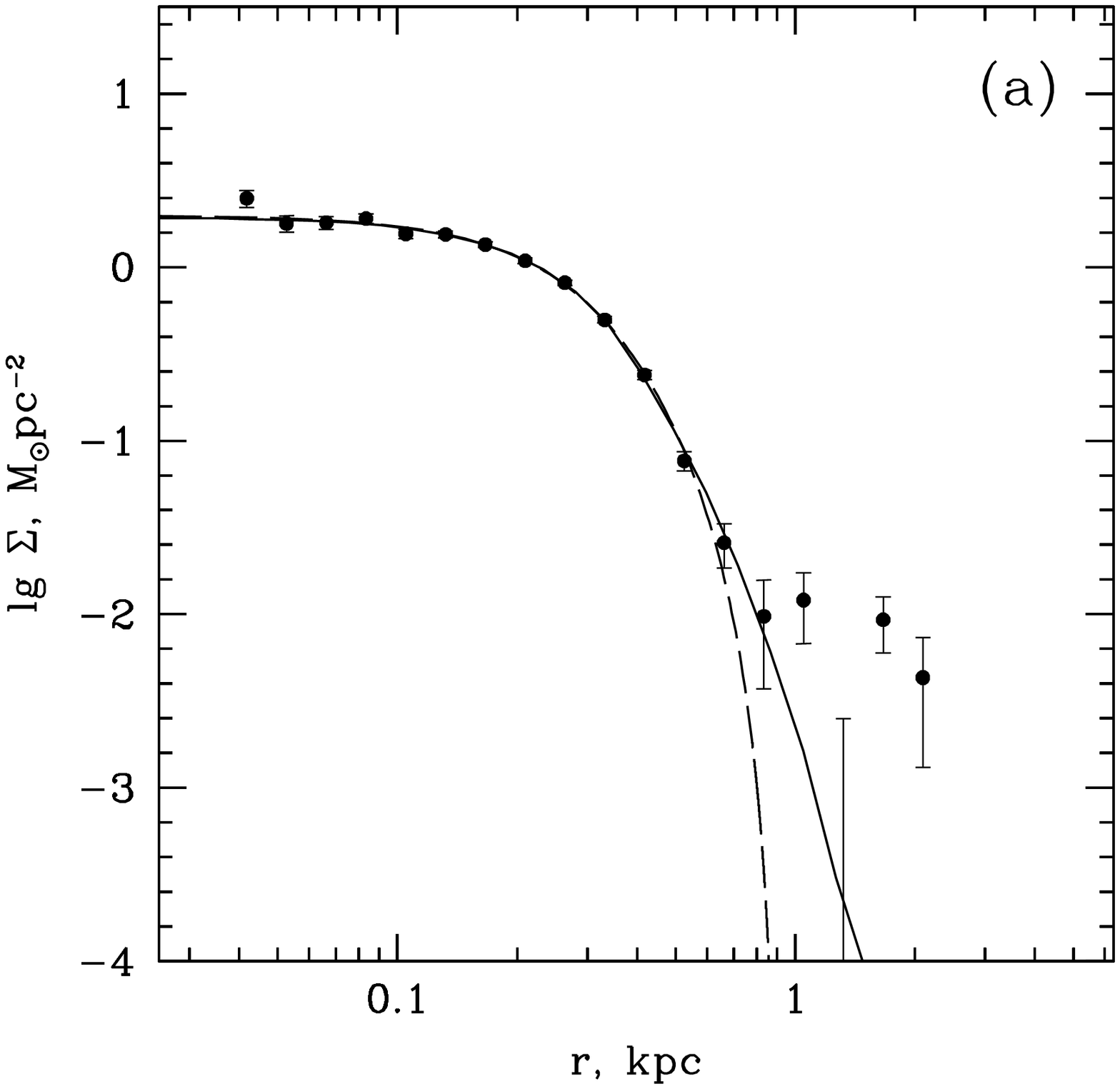}{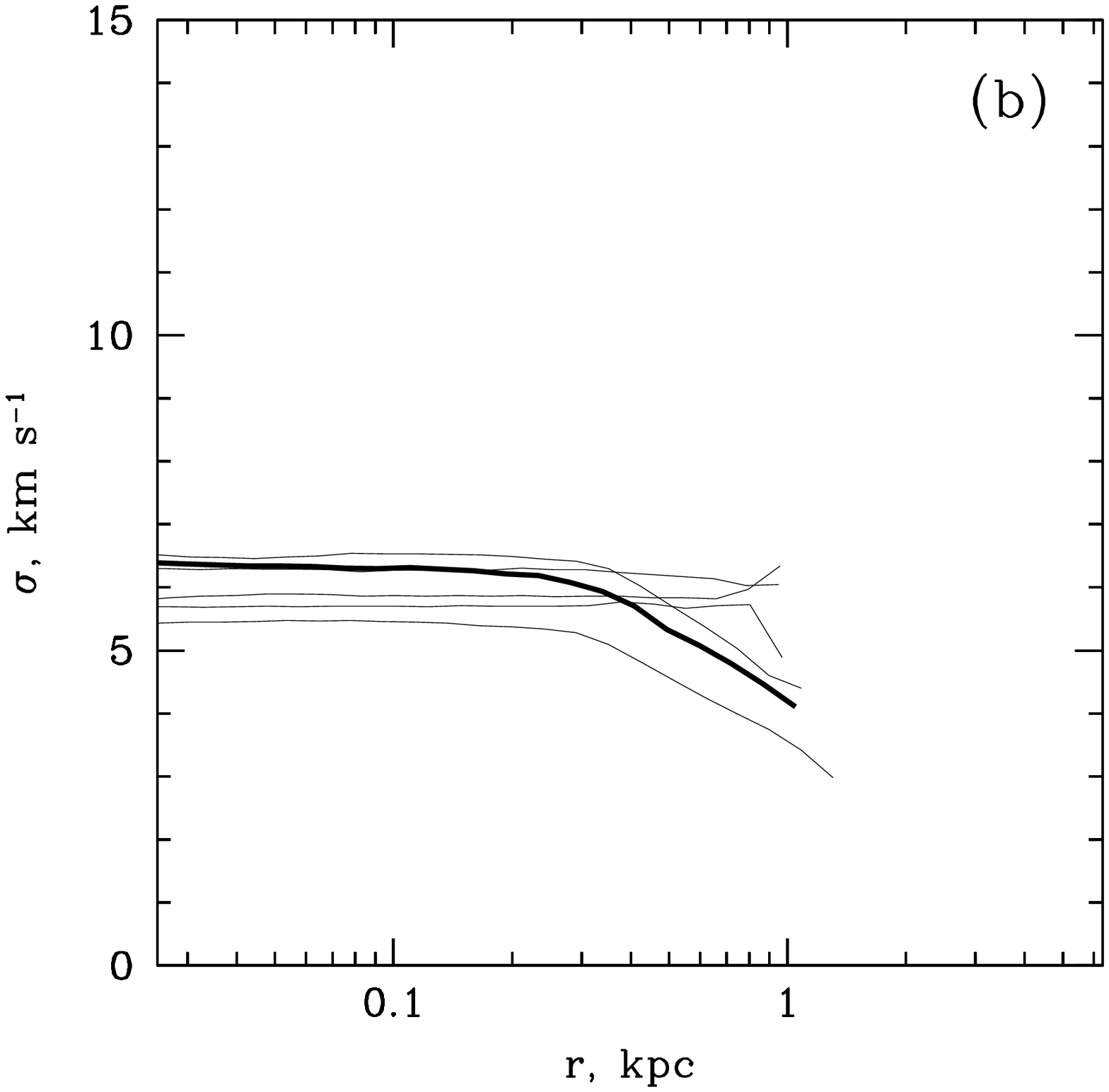}
\caption {Best fitting models for Carina. (a) Surface density profiles. The observed profile is shown
as solid circles with errorbars. The solid line corresponds to the best fitting
model (with the lowest $\chi'^2$).  The dashed line shows the best fitting
theoretical King model profile. (b) Line-of-sight velocity dispersion profiles
for all 6 ``good'' models from Table~\ref{tab2}. The thick line corresponds to
the model with the lowest $\chi'^2$.
\label{fig7} }
\end{figure*}

Similarly to Sculptor, the surface brightness profile for Carina is very noisy
at $\Sigma\lesssim 0.01\Sigma_0$ (see Figures~\ref{fig6}a and \ref{fig7}a), with
$\Sigma$ being negative at $r\sim 1.3$~kpc. We found 6 ``good'' models for
Carina, with $\chi'^2=0.84$ for the best fitting model (Table~\ref{tab2}).  As
in the cases of Draco and Sculptor, the surface brightness profiles of our Carina
models are identical to the best fitting theoretical King model in the inner
part of the galaxy, and become less steep in its outskirts.

Unlike Draco and Sculptor, all  ``good'' Carina models have Burkert DM
density profile. All virial masses are larger than $\sim 3\times
10^8$~M$_\odot$, and are typically around $10^9$~M$_\odot$. [\citet{mat98}
gives $1.3\times 10^7$~M$_\odot$.] Such masses are large enough to keep fully
photoionized ISM gravitationally bound, which is a prerequisite for the radiation
harassment picture proposed by \citet{MCB04} to explain the complex star
formation history of this dwarf \citep{mat98}. All 6 models have a halo
formation redshift of $\sim 5.7$ (lookback time 12.6~Gyr), which is consistent
with the $\sim 10-14$~Gyr age of the oldest stars in this galaxy \citep{mat98}.
The rest of the model parameters are comparable to the Draco and Sculptor cases.

\subsection{Fornax}
\label{For}

\begin{figure*}
\plottwo{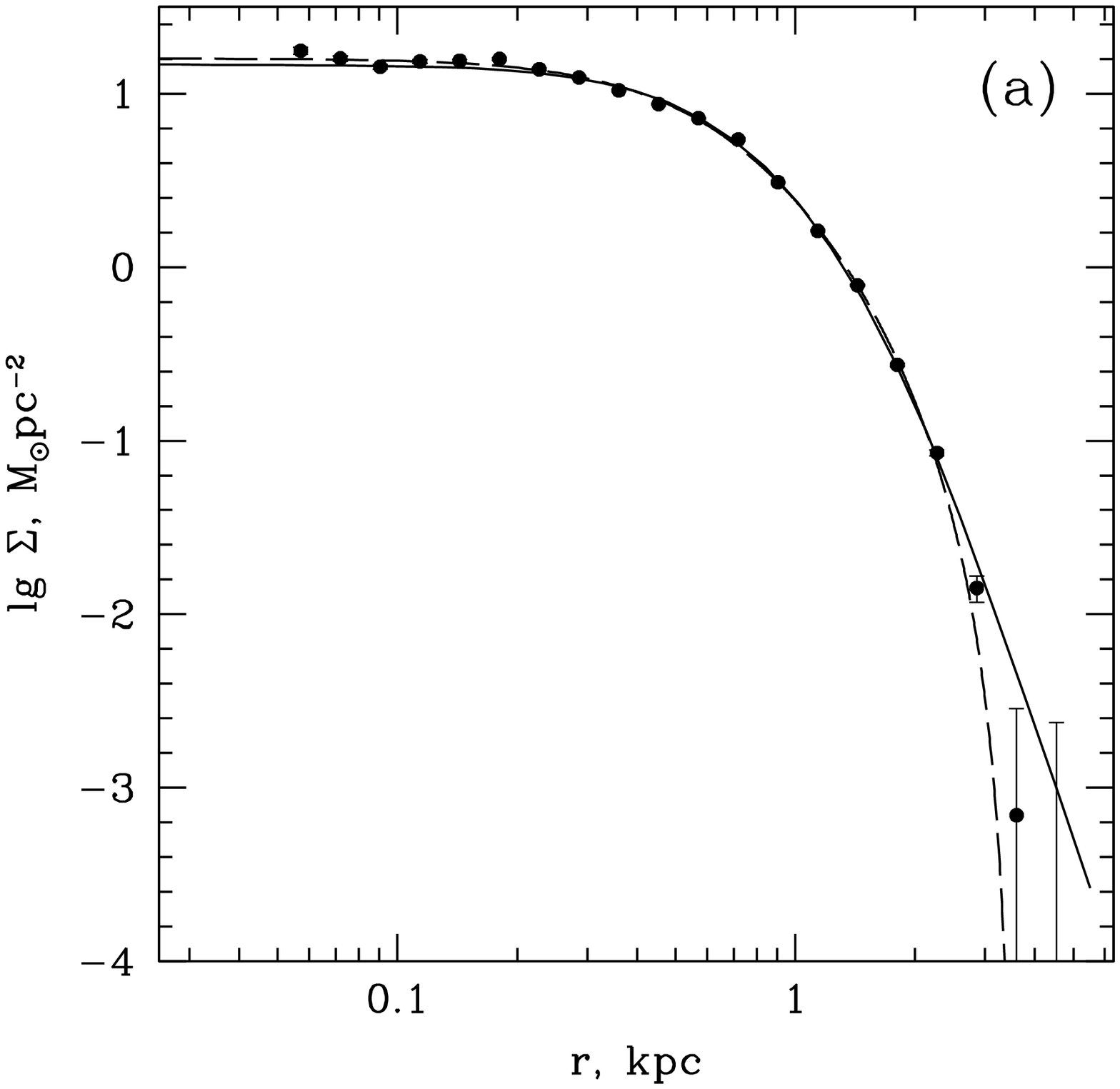}{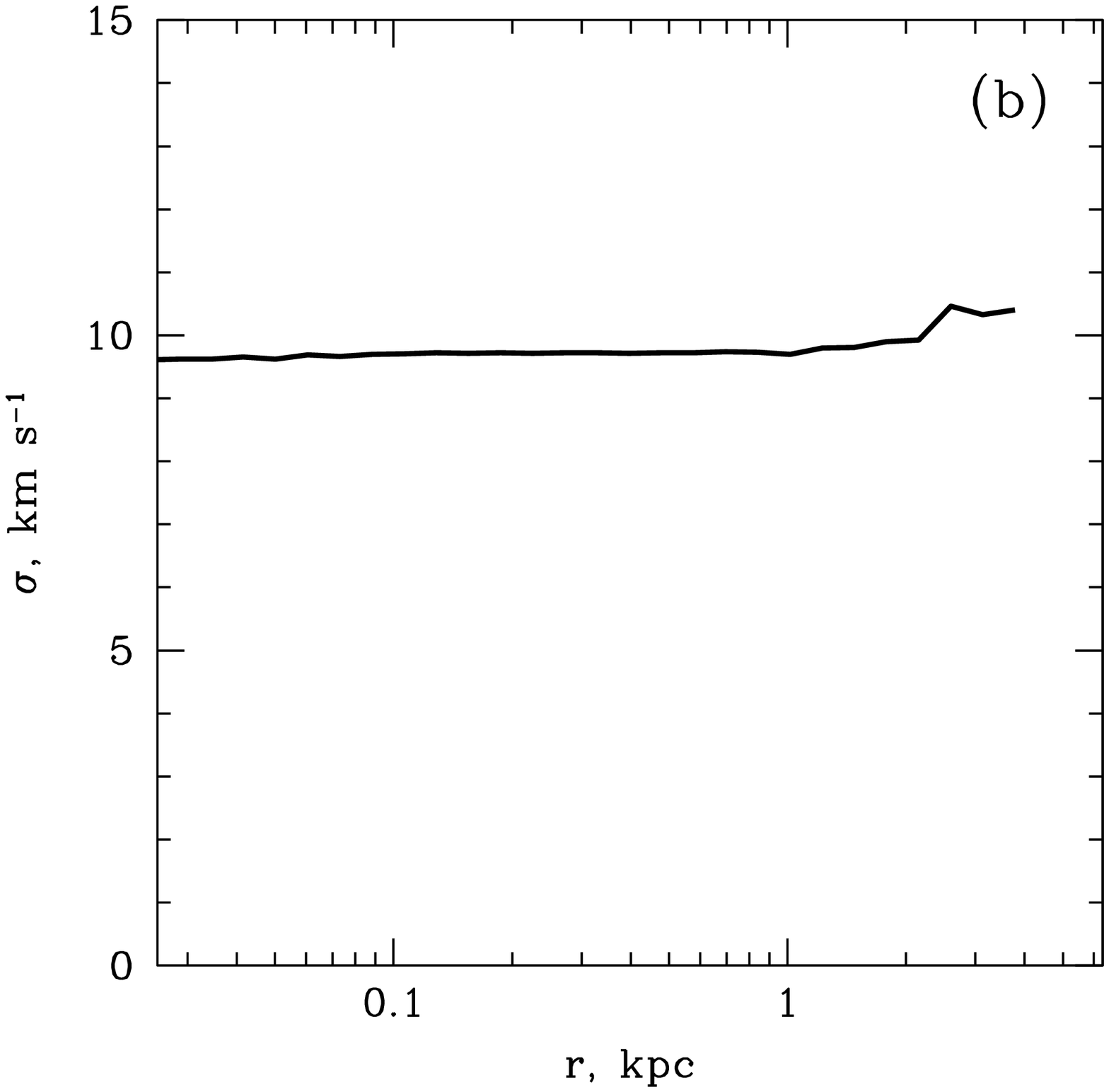}
\caption {Best fitting model for Fornax. (a) Surface density profiles. The observed profile is shown
as solid circles with errorbars. The solid line corresponds to the best fitting
model (with the lowest $\chi'^2$).  The dashed line shows the best fitting
theoretical King model profile. (b) Line-of-sight velocity dispersion profile
for the model with the lowest $\chi'^2$.
\label{fig8} }
\end{figure*}

Fornax is the only galaxy in our sample with not a single model which would
simultaneously meet all the three selection criteria from \S~\ref{par0}.  In
Table~\ref{tab2} we list parameters for the only model which has a comparable
$\chi^2$ to the best fitting King model: $\chi'^2=1.09$. The second-to-best
model has $\chi'^2=1.9$.

One could try to argue that the failure to find a model with $\chi'^2<1$ is due
to the fact that the King model already provides an almost perfect fit to the
data (see Figure~\ref{fig8}a).  It is conceivable that with unlimited computing
resources we would be able to fine-tune our initial parameters to find a model
which would fit the observed surface brightness profile slightly better than the
King model. However, King-like profiles are not natural for our models (which
typically have less steep outer density profiles), and it would be very unlikely
that it is a mere accident that Fornax happened to have an almost perfect King
profile. The tidal stripping process, on the other hand, drives naturally an
isothermal stellar system toward a King-like state, which makes it a more
attractive explanation for the Fornax observed properties.

The inferred parameters for our best fitting model (Table~\ref{tab2}) make
even more questionable the applicability of our model to Fornax. In particular,
the inferred central gas density $\rho_{g,0}\sim 0.9$~cm$^{-3}$ is much lower
than for the other three dSphs. Furthermore, all models with $\chi'^2\lesssim 3$
have $\lambda=0$. It seems unphysical for star formation to take place
with almost 100\% efficiency (see Table~\ref{tab2}) at densities $\ll
1$~cm$^{-3}$.

All of the above leads us to conclude that our model (at least in its present,
simplest form) is not applicable to the Fornax dSph. It remains to be seen if some
other non-tidal mechanism could produce naturally a King-like surface brightness
profile in a dSph. For now, Fornax being a tidally limited system seems to be
the most likely explanation. It is interesting to note that even in the tidally
limited scenario, Fornax appears to have a quite large mass, with the ``mass
follows light'' estimate of $6.8\times 10^7$~M$_\odot$ \citep{mat98}, and
potentially even larger value if Fornax is more DM dominated in its outskirts.

\section{DISCUSSION}
\label{discussion}

The observational data on three galaxies (Draco, Sculptor, and Carina) out of
four dSphs studied in this paper appear to be consistent with our simple model
of star formation in extended DM halos, and as a consequence are consistent
with them having very massive DM halos. The galaxy with the best quality star
count profiles (Draco) is also the one which makes the most convincing case for
having an extended halo --- both in the analysis of \citet{ode01} who showed
that the outer stellar isophotes for Draco are not tidally distorted, and in the
analysis carried out in this paper.

The data for Fornax appear not to be consistent with the predictions of our
model. The surface brightness profile for this galaxy is almost perfectly
described by a theoretical King model, which was derived for an isothermal
system exposed to external tidal field. This type of profile does not come
naturally in our model. In addition, the best fitting model for Fornax predicts
that the star formation in this dwarf should have taken place at unphysically
low densities of $\ll 1$~cm$^{-3}$. The results of our simulations are thus
suggestive of the cutoff in the outer surface brightness profile of Fornax
being caused by the tidal field of the Milky Way. 

The situation is significantly different with the three other dSphs we studied
in this paper (Draco, Sculptor, and Carina), which appear to have extended DM
halos. This difference can be explained if (1) the orbit of Fornax  has a significantly
smaller pericenter distance than the orbits of the other three galaxies [but
this seems to be inconsistent with the available proper motion measurements
\citep{din04}, which imply that Fornax is on a low eccentricity orbit and is
currently near its pericenter], and/or (2) the average DM density within the
stellar extent of Fornax (radius of $\sim 4$~kpc) is significantly smaller than
the average DM density within the same radius of the three other dSphs. (The
latter would be the case if Fornax was formed more recently than the three other
galaxies.)

The conclusion we reach here is quite contrary to the conventional argument that
the presence of ``extratidal'' stars (those which populate the outer halo of the
dwarf galaxy beyond the tidal radius of the best-fitting King model) around
Milky Way satellites is a good evidence for the dwarf galaxy to have undergone a
tidal disruption process, with the observed ``extratidal'' stars being part of
the halo of tidally stripped stars.  In our model, star formation in extended DM
halos naturally produces surface brightness profiles which are identical to King
profiles out to a few core radii, and are less steep in the outer parts of the
galaxy. The example of Fornax suggests that actually the absence of
``extratidal'' halo, with the galactic surface brightness profile being
described almost perfectly by theoretical King profile, could be an evidence for
the galaxy to be observed all the way to its tidal radius. In such tidally
limited systems a small number of stars do have to populate an extratidal halo,
forming leading and trailing tidal tails, but for stable satellite galaxies the
surface brightness of such tails can be very low (especially if dSphs have more
DM in their outskirts than at the center, which appears to be the case for
Draco), potentially below the sensitivity of existing observations. The systems
where tidal tails have been reliably detected (such as Sagittarius dSph and
Palomar~5) are at the last stage of being tidally disrupted, which cannot be the
case for most Galactic dSphs. We argue that to rule out our alternative,
non-tidal explanation for the presence of ``extratidal'' stars around some
Galactic dSphs, unambiguous observational evidence for these stars forming
leading and trailing tidal tails has to be obtained.

Recent observations suggest that in some Galactic dSphs (Draco, Ursa Minor, and
Sextans) the very outskirts of the galaxies have lower values of line-of-sight
velocity dispersion than the bulk of the galaxy \citep{wil04,kle04}. In our
models, this is the case for all models with sufficiently large lower density
cutoffs for star formation: $\lambda\gtrsim 0.1$ (see
Figures~\ref{fig4}b-\ref{fig6}b). For such models, we observe the line-of-sight
velocity dispersion to decline by a factor of two in the outer parts of the
galaxy, which is caused by strong radial anisotropy of stars which were born
within the initial stellar cluster radius $r_\lambda$ and then later populated
the $r>r_\lambda$ parts of the galaxy.

The Fornax \citep{mat97}, Draco \citep{wil04}, and Sextans \citep{kle04}
observations suggest that these galaxies are kinematically colder at the center
than in the intermediate regions. Such behavior cannot be explained in terms of
our single star burst model. Many dSphs however have complex ancient ($\gtrsim
10$~Gyr old) star formation histories, with slightly younger and/or more metal
rich stars being more concentrated toward the center \citep{har01,tol04}. A
natural extension of our model then would be to assume that a fraction of gas,
which was somewhat metal-enriched and expelled to large radii during the first
star burst, was later reaccreted, leading to the second episode of star
formation.  The reaccretion of the gas and the second star burst can be delayed
by up to a few Gyrs by supernovae type Ia occurring in the original stellar
population \citep{bur97}. Let us consider the simplified case of negligible
self-gravity of gas [$m'_g\ll M'(x)$ in eq.~(\ref{ode1})], with the most of the
gas being within the scaling radius $r_s$ of the DM halo [so that $M'(x)\propto
x^\alpha$, where $\alpha=2$ for NFW halos, and 3 for Burkert halos]. With the
above simplifying assumptions, equation~(\ref{ode1}) can be rewritten as

\begin{equation}
\frac{d\rho'_g}{dx}\propto-\frac{\rho'_g}{c_g'^2}x^{\alpha-2}.
\end{equation}

\noindent The solution for this equation is

\begin{equation}
\rho_g'(x)=\rho'_{g,0} \exp\left[-\frac{Bx^{\alpha-1}}{c_g'^2}\right],
\end{equation}

\noindent where $B$ is some positive constant. For the case of $\alpha > 1$,
the total gas mass is

\begin{equation}
\label{gasmass}
m'_g\propto \rho'_{g,0} c_g'^{6/(\alpha-1)}.
\end{equation}

\noindent We assume, that a star burst in a slowly cooling and contracting gas
cloud will take place when the central gas density $\rho'_{g,0}$ reaches a
certain critical value, which is the same for the original star burst (the total
gas mass $m'_{g,1}$) and the subsequent star burst (the total gas mass
$m'_{g,2}$). Using equation~(\ref{gasmass}), we can then relate the gas parameters
for the two star bursts through

\begin{equation}
\label{sound}
\frac{c_{g,2}'}{c_{g,1}'} = \left(\frac{m'_{g,2}}{m'_{g,1}}\right)^{(\alpha-1)/6}.
\end{equation}

\noindent Here $c_{g,1}'$ and $c_{g,2}'$ are the sound speed in the gas in the first
and second star bursts. The exponent in equation~(\ref{sound}) is positive for
both Burkert and NFW halos (the exponent values are $1/3$ and $1/6$,
respectively). Due to gas consumption by star formation, and potentially some
additional gas losses into intergalactic space, the total gas mass in the second
star burst is smaller than in the original star burst: $m'_{g,2}<m'_{g,1}$.
From equation~(\ref{sound}), this results in the sound speed (and hence the
temperature) of the gas in the second star burst being lower than in the first
one: $c_{g,2}'<c_{g,1}'$. Both lower temperature and lower total gas mass will
also result in the gas cloud being more compact in the second star burst.  As a
result, a kinematically colder stellar core, slightly younger and slightly more
metal rich than the bulk of the galaxy, could be formed. A multiple star bursts
scenario can explain the radial population gradients observed in many Local
Group dSphs \citep{har01}, and possibly the kink observed in the surface
brightness profile of Draco at the radius of $\sim 25$~arcmin
\citep{wil04}. [Note that no kink is observed in the Draco profile of
\citet{ode01}.]

For the three galaxies, which appear to be consistent with our model (Draco,
Sculptor, and Carina), the values of the inferred model parameters presented in
Table~\ref{tab2} do not differ much from one galaxy to another. The ``good''
models for these galaxies have a shallow inner DM density profile (with the slope
$\gamma\sim -0.5\dots 0$), which is consistent with the observations of dwarf
spiral galaxies \citep{bur95}, dwarf irregular galaxies \citep{cot00}, and low
surface brightness galaxies
\citep[e.g.][]{mar02}. The typical virial masses for the DM halos are of
order of $\sim 10^9$~M$_\odot$, with the total range spanning from 
$9\times 10^7-2\times 10^{10}$~M$_\odot$. These masses are much larger than
those obtained under assumption of constant mass-to-light ratio, and are
consistent with the idea that dSphs represent the largest subhalos expected
to populate the Milky Way halo in cosmological $\Lambda$CDM simulations
\citep{sto02,hay03}, alleviating the ``missing satellites'' problem
of such cosmologies \citep{moo99}. The inferred formation redshifts for these
galaxies are $\sim 2-6$ (lookback time $\sim 10-13$~Gyr), which is consistent
with the age of the oldest stars in these galaxies \citep{mat98}.  The virial
radii for the DM halos at the time of their formation range from $1.4-30$~kpc,
whereas the tidal radii at their current distance from the Milky Way center are
$5-32$~kpc (typically $\sim 10$~kpc, which would correspond to $\sim 6$ angular
degrees in the sky for these three dSphs). Both virial and tidal radii are
significantly larger than the observed extent of these galaxies.

The star forming gas in the ``good'' models has a temperature of $\sim
10,000-20,000$~K if it is neutral ($\sim 5000-10,000$~K if it is ionized), a
central density of $\sim 10$~cm$^{-3}$, and a central pressure of $\sim
10^5$~K~cm$^{-3}$. The temperature of the gas is a factor of $\sim 2.5$ lower
than the virial temperature of the DM halo (defined in \S~\ref{numer}),
suggesting that the gas accreted by the DM halo experienced only modest level of
radiative cooling globally prior to the starburst. The lower density cutoff for
star formation is of order of $\sim 1$~cm$^{-3}$. The inferred star formation
efficiency in the star-forming central part of ISM is $\sim 10$\%.

We argue that our results are not very sensitive to the model assumption that
the DM potential is static. During the formation of a dwarf spheroidal galaxy,
gradual collapse of the gas cloud due to radiative cooling and gravitational
instability will lead to the adiabatic contraction of the DM halo in the central
region of the galaxy, where the enclosed gas mass becomes larger than that of
DM. Proper modeling of this process requires adding many physical processes
(live DM halo, non-equilibrium chemistry, radiative heating and cooling) with
many associated free parameters, which is beyond the scope of this
paper. Instead, we argue that our main results should not be significantly
affected by the above effect by noting that (1) in our best fitting models, gas
does not significantly dominate DM at the center of the galaxy (with the maximum
ratio of the enclosed gas mass to that of DM of 10 for the Sculptor best fitting
model; in case of the Draco1 dataset, this ratio is less than $1/3$ at any
radius), and (2) the initial adiabatic contraction of the DM halo will be
compensated by the adiabatic expansion caused by the removal of $\sim 90$\% of
baryons from the central part of the galaxy by stellar feedback mechanisms after
the starburst.

Another potential pitfall is our assumption that the tidal field of Milky Way
played no role in shaping the stellar outskirts of the dSphs we studied.
Judging from the data presented in Table~\ref{tab2}, this assumption appears to
be justified. In particular, the inferred tidal radii are significantly larger
than the observed extent of these dwarfs (by a factor of $>2$). We do not expect
many dSph stars to be tidally stripped under these circumstances. More accurate
analysis of this effect is complicated as it would involve high resolution
$N$-body simulations of a two-component (DM $+$ stars) dwarf satellite in the
potential of the host galaxy with some prescription for dynamical friction to
describe the orbital decay, and with many poorly known orbital and Milky Way
halo parameters. This task is beyond the scope of our paper, where our main goal
is to sample well the initial model parameter space, with the only practical way
of achieving this being in keeping the number of free parameters as small as
possible.

The selective nature of the tidal disruption of satellites in the halo of the
host galaxy could make our method of finding the most likely DM halo progenitors
for Galactic dSphs (\S~\ref{par3}) invalid. However, this effect would be
important only if the DM halos in our best fitting models were relatively easy
to disrupt over the Hubble time in the tidal field of Milky Way. In the case of
the Draco1 dataset, the best fitting model has an NFW DM density profile and the
inferred tidal radius of $\sim 4.4 r_s$ (Table~\ref{tab2}). Numerical
simulations of \citet{hay03} demonstrated that NFW subhalos
orbiting in the host halo potential stay relatively intact after $\gtrsim 10$
orbits if $r_{\rm tid}>2r_s$. In this respect, the Draco1 model appears to be
stable. In the three remaining datasets (Draco2, Sculptor, and Carina; we
exclude Fornax from this analysis), the DM halos of the best fitting models have
Burkert profiles, with the tidal radii $r_{\rm tid}>10r_s$.  Burkert satellites
are easier to disrupt tidally than NFW satellites, as their binding radius is
$\sim 2.1$ times larger \citep{mas05b}. Nevertheless, the large inferred value
of the tidal radius ($>10r_s$) makes it very unlikely that the Burkert DM halos
corresponding to our best fitting models would be completely destroyed by the
tidal field of Milky Way in a Hubble time.

We believe that our model is consistent with the observed metallicities of
Galactic dSphs.  As can be seen in Fig.~7 of \citet{mat98} and Fig.~1 of
\citet{tam01}, the low-luminosity objects we are interested in show a relatively
large spread in metallicity (with a typical value of [Fe/H]$\sim -1.8$) and no
statistically significant correlation between metallicity and luminosity or
(conventionally derived) virial mass. The only exception among the Galactic
dSphs is Fornax, which is both the brightest and the most metal-rich
([Fe/H]$\sim -1.3$, \citealt{mat98}). As we discussed above, Fornax is not well
described by our model. If we relax the single starburst assumption, more
massive galaxies can become more metal-rich over time, as they have deeper
gravitational potential wells and larger tidal radii, enabling them to reaccrete
more metal-enriched gas expelled in the previous episode of star formation. Our
model is consistent with the data if the Galactic dSphs were formed from a
pre-enriched intergalactic medium, with [Fe/H]$\sim -2$.

\section{CONCLUSIONS}
\label{conclusions}

Our simple model of stars forming in an extended DM halo from isothermal gas,
and later dynamically relaxing after expelling the leftover gas, is consistent
with the observed properties of three out of four dSphs studied in this
paper.

The results of our simulations suggest that there is an alternative, non-tidal
explanation for the observed presence of halos of ``extratidal'' stars around
some Galactic dSphs. In our model, such halos are formed when a freshly formed
stellar cluster is relaxed dynamically inside the extended DM halo.

The virial masses of DM halos, inferred from fitting our models to the three
dSphs (Draco, Sculptor, and Carina), are in the right range ($\sim
10^9$~M$_\odot$) to give support to the idea that the Galactic dSphs represent
the most massive subhalos predicted to orbit in the Milky Way halo by
cosmological $N$-body $\Lambda$CDM simulations, alleviating in this way the
``missing satellites'' problem. The masses are also large enough to keep a fully
ionized ISM gravitationally bound, which is an essential ingredient of some
models \citep{bur97,MCB04} proposed to explain the complex star formation
history of some Galactic dSphs.

\acknowledgements
We would like to thank Mark Wilkinson and Carl Walcher for providing the star
count profiles for the dSphs, and Dean McLaughlin for making his set of
theoretical King models available to us. The $N$-body simulations reported in
this paper were carried out at the Canadian Institute for Theoretical
Astrophysics.

\begin{deluxetable}{llcccccccccccccccccccc}
\tablecaption{Results of simulations\label{tab2}} 
\tabletypesize{\scriptsize}
\tablehead{
\colhead{Parameter\rule{0pt}{5mm}}                &\colhead{Unit}&\colhead{Statistics}&\colhead{Draco1}    & \colhead{Draco2} & \colhead{Sculptor\tablenotemark{a}}& \colhead{Carina}   & \colhead{Fornax\tablenotemark{a}}
}
\startdata
$N_g$\rule{0pt}{5mm}\strut                              &              &\nodata             & 11                 & 13                 & 3                  & 6                  & 1                 \\
\tableline																														                                        
$\chi'^2$                          &              & best               & $ 0.67$            & $ 0.58$            &$ 0.94$            & $ 0.84$            &$ 1.09$            \\ 
                                   &              & mean               & $ 0.78\pm  0.11$   & $ 0.76\pm  0.14$   &$ 1.03\pm  0.08$   & $ 0.93\pm  0.06$   &  \nodata          \\ 
                                   &              & range              & $ 0.67\dots  0.97$ & $ 0.58\dots  1.00$ &$ 0.94\dots  1.09$ & $ 0.84\dots  1.00$ &  \nodata          \\ 
\tableline
$\lg F'$                           &              & best               & $ 0.70$            & $ 1.44$            &$ 1.89$            & $ 1.91$            &$ 1.30$            \\
                                   &              & mean               & $ 0.70\pm  0.34$   & $ 1.27\pm  0.69$   &$ 0.79\pm  0.95$   & $ 1.18\pm  0.74$   &  \nodata          \\ 
                                   &              & range              & $ 0.02\dots  1.31$ & $ 0.09\dots  2.21$ &$ 0.12\dots  1.89$ & $ 0.05\dots  2.07$ &  \nodata          \\ 
\tableline								 		      		      	  		     		    		            														                                        
$\gamma$                           &              & best               & $-1.00$            & $ 0.00$            &$ 0.00$            & $ 0.00$            &$ 0.00$            \\ 
                                   &              & mean               & $-0.82\pm  0.40$   & $-0.54\pm  0.52$   &$-0.67\pm  0.58$   & $ 0.00\pm  0.00$   &  \nodata          \\ 
                                   &              & range              & $-1.00\dots  0.00$ & $-1.00\dots  0.00$ &$-1.00\dots  0.00$ & $ 0.00\dots  0.00$ &  \nodata          \\ 
\tableline								 		      		      	  		     		    		            														                                        
$\lg c_g'$                         &              & best               & $-0.79$            & $-0.62$            &$-0.80$            & $-0.80$            &$-0.80$            \\ 
                                   &              & mean               & $-0.79\pm  0.04$   & $-0.66\pm  0.18$   &$-0.72\pm  0.06$   & $-1.05\pm  0.23$   &  \nodata          \\ 
                                   &              & range              & $-0.87\dots -0.69$ & $-1.10\dots -0.49$ &$-0.80\dots -0.69$ & $-1.40\dots -0.80$ &  \nodata          \\
\tableline								 		      		      	  		     		    		            														                                        
$\lg \rho_{g,0}'$                  &              & best               & $ 1.30$            & $ 0.48$            &$ 1.00$            & $ 0.00$            &$ 0.00$            \\ 
                                   &              & mean               & $ 0.83\pm  0.65$   & $ 0.66\pm  0.68$   &$ 1.36\pm  0.42$   & $-0.12\pm  0.44$   &  \nodata          \\ 
                                   &              & range              & $ 0.00\dots  1.79$ & $-1.00\dots  1.51$ &$ 1.00\dots  1.81$ & $-1.00\dots  0.14$ &  \nodata          \\ 
\tableline								 		      		      	  		     		    		            														                                        
$\lg \xi$                          &              & best               & $-1.00$            & $-1.29$            &$-0.63$            & $-1.00$            &$-0.15$            \\ 
                                   &              & mean               & $-0.66\pm  0.45$   & $-0.95\pm  0.25$   &$-0.37\pm  0.28$   & $-0.83\pm  0.41$   &  \nodata          \\ 
                                   &              & range              & $-1.10\dots  0.00$ & $-1.29\dots -0.41$ &$-0.63\dots -0.08$ & $-1.00\dots  0.00$ &  \nodata          \\ 
\tableline								 		      		      	  		     		    		            														                                        
$\lambda$                          &              & best               & $ 0.10$            & $ 0.10$            &$ 0.00$            & $ 0.10$            &$ 0.00$            \\ 
                                   &              & mean               & $ 0.14\pm  0.08$   & $ 0.15\pm  0.11$   &$ 0.07\pm  0.06$   & $ 0.08\pm  0.12$   &  \nodata          \\ 
                                   &              & range              & $ 0.10\dots  0.30$ & $ 0.00\dots  0.30$ &$ 0.00\dots  0.10$ & $ 0.00\dots  0.30$ &  \nodata          \\ 
\tableline								 		      		      	  		     		    		            														                                        
$\lg M_s$                          &M$_\odot$     & best               & $ 9.75$            & $ 8.48$            &$ 8.59$            & $ 8.59$            &$ 9.25$            \\ 
                                   &              & mean               & $ 9.47\pm  0.33$   & $ 8.86\pm  0.70$   &$ 8.86\pm  0.26$   & $ 9.26\pm  0.65$   &  \nodata          \\ 
                                   &              & range              & $ 8.88\dots 10.05$ & $ 7.98\dots  9.96$ &$ 8.59\dots  9.12$ & $ 8.49\dots 10.26$ &  \nodata          \\       
\tableline
$\lg r_s$                          &kpc           & best               & $ 0.54$            & $-0.30$            &$-0.13$            & $-0.12$            &$ 0.27$            \\ 
                                   &              & mean               & $ 0.41\pm  0.27$   & $ 0.02\pm  0.42$   &$ 0.23\pm  0.32$   & $ 0.14\pm  0.23$   &  \nodata          \\ 
                                   &              & range              & $-0.10\dots  0.74$ & $-0.56\dots  0.59$ &$-0.13\dots  0.47$ & $-0.12\dots  0.49$ &  \nodata          \\ 
\tableline								 		      		          		     		    		            																		                                        
$\lg M_{\rm vir}$                  &M$_\odot$     & best               & $ 9.79$            & $ 8.45$            &$ 8.54$            & $ 8.55$            &$ 9.27$            \\ 
                                   &              & mean               & $ 9.49\pm  0.34$   & $ 8.86\pm  0.72$   &$ 8.84\pm  0.28$   & $ 9.23\pm  0.66$   &  \nodata          \\ 
                                   &              & range              & $ 8.84\dots 10.06$ & $ 7.93\dots  9.95$ &$ 8.54\dots  9.11$ & $ 8.46\dots 10.24$ &  \nodata          \\ 
\tableline								 		      		          		     		    		            																		                                        
$a$                                &              & best               & $ 0.35$            & $ 0.10$            &$ 0.14$            & $ 0.14$            &$ 0.24$            \\ 
                                   &              & mean               & $ 0.34\pm  0.11$   & $ 0.22\pm  0.13$   &$ 0.33\pm  0.17$   & $ 0.15\pm  0.01$   &  \nodata          \\ 
                                   &              & range              & $ 0.11\dots  0.42$ & $ 0.09\dots  0.40$ &$ 0.14\dots  0.43$ & $ 0.14\dots  0.17$ &  \nodata          \\ 
\tableline								 		      		          		     		    		            																		                                        
$\nu_c$                            &              & best               & $-0.94$            & $ 1.16$            &$ 0.22$            & $ 0.14$            &$-0.56$            \\ 
                                   &              & mean               & $-0.90\pm  1.05$   & $-0.01\pm  1.23$   &$-1.60\pm  1.58$   & $ 0.30\pm  0.34$   &  \nodata          \\ 
                                   &              & range              & $-1.68\dots  1.42$ & $-1.60\dots  1.70$ &$-2.52\dots  0.22$ & $-0.22\dots  0.74$ &  \nodata          \\ 
\tableline
$\lg (\sigma_{\rm ini,0}/\sigma_0)$&              & best               & $ 0.09$            & $ 0.11$            &$ 0.08$            & $ 0.07$            &$ 0.02$            \\ 
                                   &              & mean               & $ 0.08\pm  0.01$   & $ 0.10\pm  0.02$   &$ 0.09\pm  0.02$   & $ 0.05\pm  0.02$   &  \nodata          \\ 
                                   &              & range              & $ 0.06\dots  0.10$ & $ 0.07\dots  0.14$ &$ 0.08\dots  0.12$ & $ 0.03\dots  0.08$ &  \nodata          \\ 
\tableline								 		      		          		     		    		            																		                                        
$\lg T$                            &K             & best               & $ 4.43$            & $ 4.35$            &$ 3.93$            & $ 3.93$            &$ 4.18$            \\ 
                                   &              & mean               & $ 4.28\pm  0.09$   & $ 4.32\pm  0.10$   &$ 3.98\pm  0.09$   & $ 3.84\pm  0.08$   &  \nodata          \\ 
                                   &              & range              & $ 4.12\dots  4.43$ & $ 4.11\dots  4.48$ &$ 3.93\dots  4.08$ & $ 3.76\dots  3.93$ &  \nodata          \\ 
\tableline								 		      		          		     		    		            																		                                        
$\lg \rho_{g,0}$                   &cm$^{-3}$     & best               & $ 0.87$            & $ 1.29$            &$ 1.41$            & $ 0.38$            &$-0.16$            \\ 
                                   &              & mean               & $ 0.50\pm  0.50$   & $ 0.86\pm  0.34$   &$ 0.94\pm  0.48$   & $ 0.16\pm  0.46$   &  \nodata          \\ 
                                   &              & range              & $-0.23\dots  1.04$ & $ 0.10\dots  1.29$ &$ 0.45\dots  1.41$ & $-0.77\dots  0.38$ &  \nodata          \\ 
\tableline								 		      		          		     		    		            																		                                        
$\lg P$                            &K~cm$^{-3}$   & best               & $ 5.30$            & $ 5.63$            &$ 5.35$            & $ 4.31$            &$ 4.03$            \\ 
                                   &              & mean               & $ 4.77\pm  0.57$   & $ 5.18\pm  0.36$   &$ 4.92\pm  0.49$   & $ 3.99\pm  0.48$   &  \nodata          \\ 
                                   &              & range              & $ 3.89\dots  5.41$ & $ 4.48\dots  5.69$ &$ 4.39\dots  5.35$ & $ 3.03\dots  4.31$ &  \nodata          \\ 
\tableline								 		      		          		     		    		            																		                                        
$\rho_\lambda$                     &cm$^{-3}$     & best               & $ 0.74$            & $ 1.94$            &$ 0.00$            & $ 0.24$            &$ 0.00$            \\
                                   &              & mean               & $ 0.66\pm  0.54$   & $ 1.16\pm  0.90$   &$ 0.39\pm  0.46$   & $ 0.08\pm  0.11$   &  \nodata          \\ 
                                   &              & range              & $ 0.06\dots  1.72$ & $ 0.00\dots  2.68$ &$ 0.00\dots  0.90$ & $ 0.00\dots  0.24$ &  \nodata          \\ 
\tableline								 		      		          		     		    		            																		                                        
$\lg r_{\rm vir}$                  &kpc           & best               & $ 1.31$            & $ 0.35$            &$ 0.50$            & $ 0.51$            &$ 0.99$            \\ 
                                   &              & mean               & $ 1.16\pm  0.31$   & $ 0.73\pm  0.47$   &$ 0.92\pm  0.37$   & $ 0.78\pm  0.23$   &  \nodata          \\ 
                                   &              & range              & $ 0.54\dots  1.47$ & $ 0.14\dots  1.36$ &$ 0.50\dots  1.17$ & $ 0.51\dots  1.16$ &  \nodata          \\ 
\tableline								 		      		          		     		    		            																		                                        
$\lg r_{\rm tid}$                  &kpc           & best               & $ 1.18$            & $ 0.88$            &$ 0.88$            & $ 0.96$            &$ 1.25$            \\ 
                                   &              & mean               & $ 1.10\pm  0.07$   & $ 0.95\pm  0.19$   &$ 0.87\pm  0.04$   & $ 1.18\pm  0.22$   &  \nodata          \\ 
                                   &              & range              & $ 1.00\dots  1.24$ & $ 0.71\dots  1.35$ &$ 0.83\dots  0.90$ & $ 0.91\dots  1.50$ &  \nodata          \\ 
\enddata
\tablecomments{Here $N_g$ is the number of ``good'' models (satisfying all three selection creteria from \S~\ref{par0}), $\chi'^2\equiv \chi^2/\chi^2_{\rm King}$, $F'\equiv F/F_{\rm min}$,
$a=1/(1+z)$, $\rho_{g,0}$ is the central number density of gas, $P=\rho_{g,0} T$ is the central gas pressure,
$\rho_\lambda=\rho_{g,0} \lambda$, $r_{\rm vir}$ is the halo virial radius at the
formation epoch, and $r_{\rm tid}$ is the current tidal radius (see definition
in text).}
\tablenotetext{a}{Models with $\chi'^2<1.1$.}
\end{deluxetable}

\end{document}